\documentclass[twocolumn,prb,citeautoscript,superscriptaddress,8pt]{revtex4-2}

\def\invs{s$^{-1}$}
\usepackage{graphicx}
\usepackage{gensymb}
\usepackage{color}
\usepackage[dvipsnames]{xcolor}
\usepackage{float}
\usepackage{upgreek}
\usepackage{bm}
\usepackage{amsmath,amssymb}
\usepackage{array}
\usepackage{hyperref}
\usepackage{multirow}
\newcolumntype{M}[1]{>{\centering\arraybackslash}m{#1}}
\newcommand{\ket}[1]{\ensuremath{\left| #1 \right\rangle}}
\newcommand{\bra}[1]{\ensuremath{\left\langle #1 \right|}}

\hypersetup{urlcolor=teal, colorlinks=true, linkcolor=teal, citecolor=teal, linktocpage=true}

\begin{document}

\title{A charge transfer mechanism for optically addressable solid-state spin pairs}

\author{Islay~O.~Robertson}
\affiliation{School of Science, RMIT University, Melbourne, VIC 3001, Australia}

\author{Benjamin~Whitefield}
\affiliation{School of Mathematical and Physical Sciences, University of Technology Sydney, Ultimo, NSW 2007, Australia}
\affiliation{ARC Centre of Excellence for Transformative Meta-Optical Systems, Faculty of Science, University of Technology Sydney, Ultimo, NSW 2007, Australia}

\author{Sam~C.~Scholten}
\affiliation{School of Science, RMIT University, Melbourne, VIC 3001, Australia}

\author{Priya~Singh}
\affiliation{School of Science, RMIT University, Melbourne, VIC 3001, Australia}

\author{Alexander~J.~Healey}
\affiliation{School of Science, RMIT University, Melbourne, VIC 3001, Australia}

\author{Philipp~Reineck}
\affiliation{School of Science, RMIT University, Melbourne, VIC 3001, Australia}

\author{Mehran~Kianinia} 
\affiliation{School of Mathematical and Physical Sciences, University of Technology Sydney, Ultimo, NSW 2007, Australia}
\affiliation{ARC Centre of Excellence for Transformative Meta-Optical Systems, Faculty of Science, University of Technology Sydney, Ultimo, NSW 2007, Australia}

\author{Gergely~Barcza} 
\affiliation{HUN-REN Wigner Research Centre for Physics, P.O.\ Box 49, H-1525 Budapest, Hungary}
\affiliation{MTA–ELTE Lend\"{u}let ``Momentum" NewQubit Research Group, P\'azm\'any P\'eter, S\'et\'any 1/A, 1117 Budapest, Hungary}

\author{Viktor~Ivády} 
\affiliation{Department of Physics of Complex Systems, E\"otv\"os Lor\'and University, Egyetem t\'er 1-3, H-1053 Budapest, Hungary}
\affiliation{MTA–ELTE Lend\"{u}let ``Momentum" NewQubit Research Group, P\'azm\'any P\'eter, S\'et\'any 1/A, 1117 Budapest, Hungary}

\author{David~A.~Broadway}
\affiliation{School of Science, RMIT University, Melbourne, VIC 3001, Australia}

\author{Igor~Aharonovich}
\email{igor.aharonovich@uts.edu.au}
\affiliation{School of Mathematical and Physical Sciences, University of Technology Sydney, Ultimo, NSW 2007, Australia}
\affiliation{ARC Centre of Excellence for Transformative Meta-Optical Systems, Faculty of Science, University of Technology Sydney, Ultimo, NSW 2007, Australia}

\author{Jean-Philippe~Tetienne}
\email{jean-philippe.tetienne@rmit.edu.au}
\affiliation{School of Science, RMIT University, Melbourne, VIC 3001, Australia}

\begin{abstract}
Optically detected magnetic resonance (ODMR) with no resolvable zero-field splitting has been observed from emitters in hexagonal boron nitride across a broad range of wavelengths, but so far an understanding of their microscopic structure and the physical origin of ODMR has been lacking. 
Here we perform comprehensive measurements and modelling of the spin-resolved photodynamics of ensembles and single emitters, and uncover a universal model that accounts, and provides an intuitive physical explanation, for all key experimental features.
The model, inspired by the radical-pair mechanism from spin chemistry, assumes a pair of nearby point defects -- a primary optically active defect and a secondary defect. Charge transfer between the two defects creates a metastable weakly coupled spin pair with ODMR naturally arising from selection rules. Using first-principle calculations, we show that simple defect pairs made of common carbon defects provide a plausible microscopic explanation.    
Our optical-spin defect pair (OSDP) model resolves several previously open questions including the asymmetric envelope of the Rabi oscillations, the large variability in ODMR contrast amplitude and sign, and the wide spread in emission wavelength. 
It may also explain similar phenomena observed in other wide bandgap semiconductors such as GaN. 
The presented framework will be instrumental in guiding future theoretical and experimental efforts to study and engineer solid-state spin pairs.
\end{abstract}

\maketitle

In the last few years, hexagonal boron nitride (hBN) has garnered significant interest as a host for optically addressable spins, motivated by the unique attributes afforded by the material's layered van der Waals structure~\cite{GottschollNatMat2020, GottschollSciAdv2021, Liu2022NatComm, GaoNatMat2022, GongNatComms2023, HealeyNP2022, Ramsay2023, Rizzato2023}, differentiating it from established hosts such as diamond and silicon carbide~\cite{Atature2018,Wolfowicz2021}. 
In particular, the prospect of being able to engineer spin defects in ultrathin (few-layer) hBN flakes, potentially down to monolayer, is appealing for quantum sensing applications where near atomic scale proximity between sensor and target may be achievable~\cite{DurandPRL2023, RobertsonACSNano2023, GaoACSPhoto2023, Zhou2024}, and for quantum simulations where two-dimensional confinement is highly desirable~\cite{Yao2018, Davis2023}.
Currently, the most explored hBN spin defect is the negatively charged boron vacancy ($V_{\rm B}^-$) center, which emits in the near-infrared and has a spin-triplet ($S=1$) ground state with optically detected magnetic resonance (ODMR) enabled via a spin-dependent intersystem crossing \cite{GottschollNatMat2020, Ivady2020, Reimers2020, Kianinia2020, Xingyu2021, Liu2021, Mathur2022, haykalDecoherenceVBSpin2022}. 
So far the $V_{\rm B}^-$ defect has only been observed in ensembles and its practicality is limited owing to its low quantum efficiency~\cite{Reimers2020}, motivating the continued exploration of new spin defects.

Recently, several groups have reported on a family of substantially brighter spin defects emitting over a wide range of wavelengths, with zero-phonon lines reported from about 420\,nm to 800\,nm~\cite{MendelsonNatMat2021, ChejanovskyNatMat2021, SternNatComms2022, GuoNatComms2023, YangACSAPN2023, Scholten2023, Patel2023,Singh2024}.
Common to these defects is the absence of a distinct fine or hyperfine structure in the ODMR spectrum, which features a singular resonance 
akin to a $S=1/2$ electronic system.
Despite many attempts, there is still no experimental or theoretical consensus on the atomic and electronic structure of these spin defects~\cite{Auburger2021, Pinilla2023}. 
It has been experimentally shown carbon impurities are involved~\cite{MendelsonNatMat2021}, and many substitutional carbon defects have subsequently been proposed as potential candidates including monomers, dimers, trimers, tetramers, as well as pairs of non-adjacent point defects~\cite{MendelsonNatMat2021, ChejanovskyNatMat2021, GuoNatComms2023, Auburger2021, Golami2022, Tan2022, Benedek2023, Pinilla2023}. 
Most studies so far have employed density functional theory (DFT) calculations to find a candidate matching the experimental photoluminescence (PL) spectrum and exhibiting a paramagnetic ground state with a narrow hyperfine structure ($<50$\,MHz as bounded by experiments~\cite{ChejanovskyNatMat2021, SternNatComms2022, GuoNatComms2023, YangACSAPN2023, Scholten2023, Patel2023}). 
However, there has been little discussion of the origin of the ODMR response, which requires a mechanism for photo-induced spin polarisation and spin-dependent PL. 
As a result, it is unclear if any of the candidate defects proposed so far can in theory produce ODMR. 
On the other hand, several phenomenological models have been proposed to explain the spin-dependent photodynamics measured on a few single emitters~\cite{ChejanovskyNatMat2021, SternNatComms2022, Patel2023}, but it remains unclear whether the proposed models are broadly applicable beyond the handful of defects studied. 
Moreover, no attempt has been made to explain the physical origin of the spin selectivity within these models.

In this Article, we present spin- and time-resolved PL measurements averaged over a large ensemble of emitters in hBN, and propose a holistic model adapted from the established radical-pair mechanism~\cite{Steiner1989,Woodward2002,
Evans2013} which is shown to account for all key experimental observations. Furthermore, we perform first-principle calculations and identify a plausible microscopic model based on common carbon defects.
Our optical-spin defect pair (OSDP) model provides an intuitive explanation for the origin of ODMR in terms of selection rules, and explains the asymmetric envelope of the Rabi oscillations, the large variability in ODMR contrast amplitude and sign, and the wide spread in emission wavelength. 
Fundamental outstanding questions in the hBN community are resolved by the OSDP model, and may also be applicable to other systems such as the recently observed spin defects in GaN~\cite{Luo2024}. 

\section{Experimental results}

\begin{figure}[tb!]
    \centering
    \includegraphics{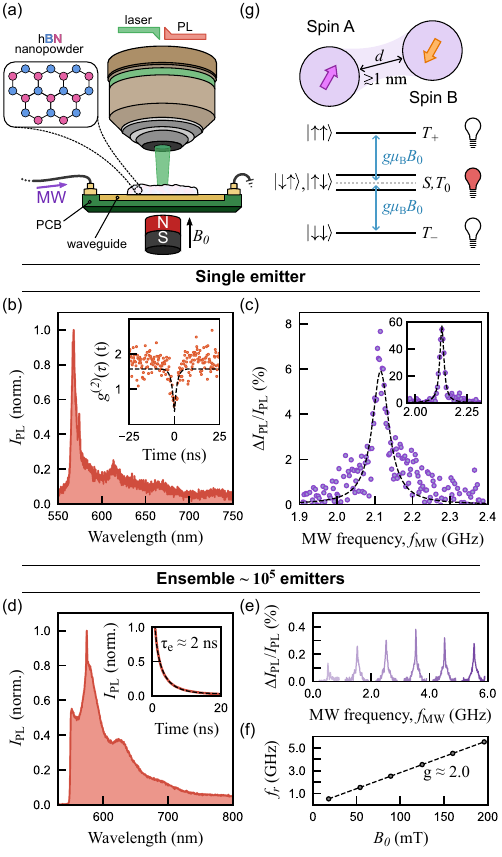}
    \caption{ \textbf{Spin-active visible-band emitters in hBN}. 
    (a)~Schematic of the experiment. 
    (b)~PL spectrum of a single emitter excited with a 532\,nm laser. 
    Inset: corresponding auto-correlation function indicating single photon emission. 
    (c)~CW ODMR spectrum of the emitter in (b), obtained at $B_0\approx75$\,mT. 
    Inset: ODMR spectrum of another emitter featuring a 60\% contrast. 
    (d)~PL spectrum of a large ensemble of emitters. 
    Inset: PL decay trace (red) following a short laser pulse, from which the excited-state lifetime $\tau_e$ is estimated via a stretched exponential fit (black dashed line). 
    (e)~Series of CW ODMR spectra of the ensemble at different field strengths $B_0\approx18 - 196$\,mT. 
    (f)~Magnetic field dependence of the resonance frequency extracted from (e). 
    Dashed line is a linear fit, indicating $g = 2.0(1)$. 
    (g)~Schematic of a weakly coupled spin pair and the resultant energy level structure. 
    The eigenstates are $T_+=\ket{\uparrow\uparrow}$, $T_-=\ket{\downarrow\downarrow}$, and mixtures of $\ket{\uparrow\downarrow}$ and $\ket{\downarrow\uparrow}$, or equivalently of $S=(\ket{\uparrow\downarrow}-\ket{\downarrow\uparrow})/\sqrt{2}$ and $T_0=(\ket{\uparrow\downarrow}+\ket{\downarrow\uparrow})/\sqrt{2}$ \cite{Boehme2003}. 
    Energy splitting between the two mixed states is assumed negligible (weakly coupled approximation).}
    \label{fig1}
\end{figure}

The experiment is depicted in Fig.~\ref{fig1}(a). 
A commercially sourced hBN nanopowder is deposited on a printed circuit board and illuminated by a 532\,nm laser beam. 
When the powder film is sufficiently sparse, single emitters can be resolved, see an example PL spectrum and auto-correlation function in Fig.~\ref{fig1}(b). 
The continuous wave (CW) ODMR spectrum for this emitter reveals a single peak at $f_r\approx g\mu_{\rm B} B_0/h$ with a positive contrast of 6\% [Fig.~\ref{fig1}(c)], where $B_0$ is the applied magnetic field (here $B_0\approx75$\,mT), $\mu_{\rm B}$ is the Bohr magneton, $h$ is Planck's constant, and $g\approx2$ the Land{\'e} $g$-factor. 
The ODMR contrast observed for single emitters ranges from non measurable (below our noise floor of 0.2\%) all the way up to 60\% [see inset of Fig.~\ref{fig1}(c) and SI Sec.~\ref{sec:singles}].   
To average out defect-to-defect variations and orientation dependence, a large ensemble of emitters in a thick powder film was measured, with an estimated $\sim 10^5$ emitters in the probed volume. 
The ensemble-averaged PL spectrum shows primary features lying between $550$ and $700$\,nm, with an average excited state lifetime of $\tau_e \approx 2$\,ns [Fig.~\ref{fig1}(d)]. 
ODMR spectra of the ensemble under a series of magnetic fields affirms the $ g=2.0(1)$ dependence, with a maximum contrast of 0.4\% [Fig.~\ref{fig1}(e,f)].

In previous work~\cite{Scholten2023}, we demonstrated that the spin system driven in ODMR is a pair of weakly coupled $S=1/2$ electron spins, as evidenced by a beat frequency emerging in the Rabi oscillations at high microwave (MW) driving power~\cite{McCamey2010,Lee2010}. 
Here, weakly coupled means spin-spin interactions (exchange and dipole-dipole) do not cause an appreciable splitting in the ODMR spectrum, constraining the distance between the two spins to $\gtrsim1$\,nm (corresponding to a dipole-dipole coupling of $\lesssim25$\,MHz), see Fig.~\ref{fig1}(g).
In this case, the four states of the spin pair can be divided into two effective states differing by their singlet ($S=0$) content: the anti-parallel spin configurations (singlet-triplet mixtures of $\ket{\uparrow\downarrow}$ and $\ket{\downarrow\uparrow}$) on the one hand, and the parallel spin configurations (pure triplet states $\ket{\uparrow\uparrow}$ and $\ket{\downarrow\downarrow}$) on the other hand. 
As we will demonstrate later in this Article, the observed ODMR contrast is a result of selection rules governing electronic transitions from/to these two effective spin states.

\begin{figure}[tb!]
    \centering
    \includegraphics{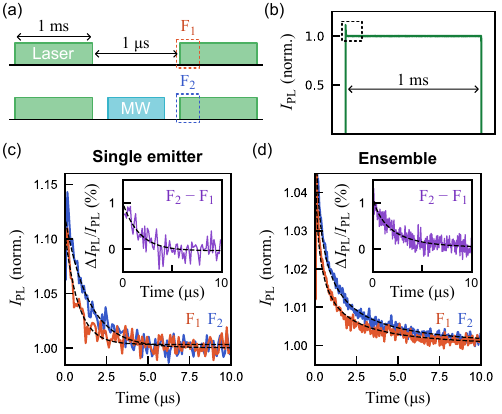}
    \caption{{\bf Spin-resolved photodynamics.} 
    (a)~Pulse sequence for the spin-resolved PL measurements. An initial 1-ms laser pulse is applied to polarise the spin. 
    The second laser pulse reads out the PL from the polarised spin following a 1-$\mu$s dark time (top sequence), or from the unpolarised spin following a 1-$\mu$s MW pulse (bottom). 
    (b)~PL trace for the entire 1-ms laser pulse. Throughout, the PL amplitude is normalised to the steady-state PL. 
    (c)~PL trace during the first $10\,\mu$s of the readout laser pulse with (blue, $F_2$) and without (orange, $F_1$) the applied MW pulse, for the same single emitter as in Fig.~\ref{fig1}(b,c). 
    Inset: Relative difference between the two traces. The dashed lines are monoexponential fits. 
    (d)~Same as (c) but for a large ensemble of emitters. Here the dashed lines are stretched exponential fits.  
    }
    \label{fig2}
\end{figure}

To gain insights into the spin-photodynamics, we recorded spin-resolved PL traces using the pulse sequences illustrated in Fig.~\ref{fig2}(a). 
A $1$-ms initialisation laser pulse is followed by a 1-$\mu$s dark time and a $1$-ms readout laser pulse, see PL trace during the readout pulse in Fig.~\ref{fig2}(b). 
For the single emitter, the PL overshoots by $\sim10\%$ as the laser is turned on, before decaying to a steady-state value after a settling time $T_{\rm sett}$ of few microseconds [Fig.~\ref{fig2}(c), orange trace]. 
This decay indicates shelving of some populations into an optically inactive state. 
When the dark time is replaced by a resonant MW pulse to mix the spin populations [Fig.~\ref{fig2}(c), blue trace], the PL overshoots to a higher value (by a few \%), consistent with the positive ODMR contrast. 
The difference between the two PL traces decays to zero in a few microseconds [Fig.~\ref{fig2}(c), inset], which is a measure of the photo-induced spin polarisation time and matches the settling time, $T_{\rm spin}\approx T_{\rm sett}$, indicating the spin polarisation dynamics is related to the shelving dynamics.
The ensemble measurement [Fig.~\ref{fig2}(d)] returns a similar behaviour to the single emitter, suggesting that the same spin-dependent photophysical processes are involved in both cases.
In the following, we will therefore focus on ensemble-averaged measurements. 

\begin{figure}[b!]
    \centering
    \includegraphics{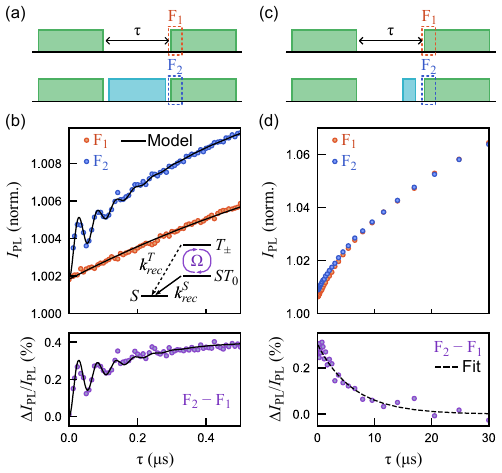}
    \caption{{\bf Optical readout of spin dynamics.} 
    (a)~Rabi measurement sequence with 5-$\mu$s gated regions at the front of the readout laser pulse following a MW pulse of variable duration $\tau$ (blue, $F_2$) or a dark time of duration $\tau$ (red, $F_1$). 
    (b)~Top: PL averaged over gated regions indicated in (a), as a function of $\tau$. 
    The solid lines are simulations based on the three-level model depicted in inset, fit to the experimental data. 
    Rabi driving at rate $\Omega$ occurs between pure triplet ($T_\pm$) and mixed singlet-triplet ($ST_0$) states which decay at different rates ($k_{\rm rec}^T$ and $k_{\rm rec}^S$ respectively) to a pure singlet ($S$) state.
    Bottom: Normalised Rabi measurement ($F_2-F_1$) and corresponding simulation.
    (c)~Sequence for the spin contrast decay measurement. 
    The readout laser pulse follows a variable dark time $\tau$ with or without a final 1-$\mu$s MW pulse. 
    (d)~Top: PL averaged over 5-$\mu$s gated regions indicated in (c), as a function of $\tau$. 
    Bottom: Difference between the two traces ($F_2-F_1$), fit with a monoexponential to estimate the average spin polarisation lifetime.
    }
    \label{fig3}
\end{figure}

We next perform a Rabi experiment.
The pulse sequence is shown in Fig.~\ref{fig3}(a), which scans the time $\tau$ between subsequent laser pulses, with ($F_2$) or without MW ($F_1$). 
We integrate over the first 5\,$\mu$s of each laser pulse (normalised to the steady-state PL), plotted against $\tau$ in Fig.~\ref{fig3}(b) (top). 
Both traces $F_1$ and $F_2$ increase with $\tau$, which is due to the recovery of the shelved populations. 
Rabi oscillations are clearly visible in the $F_2$ trace (with MW), dampening after $\tau\approx200$\,ns.
Past these oscillations, the $F_2$ trace decays faster than the $F_1$ trace. 
This difference points to a spin-dependent recovery process. 
Moreover, the highly asymmetric envelope of the Rabi oscillations even when normalised [Fig.~\ref{fig3}(b), bottom] suggests the driven spin pair is a metastable configuration \cite{KosugiPRB2005}. 
Combining these observations, we consider a three-level model where coherent spin driving at a rate $\Omega$ occurs between the two effective states of the spin pair (denoted as $T_\pm$ and $ST_0$), both of which decay incoherently to an optically active third state via the recombination rates $k_{\rm rec}^T$ and $k_{\rm rec}^S$ [Fig.~\ref{fig3}(b), inset].
This model is in excellent agreement with the data for a ratio $k_{\rm rec}^S/k_{\rm rec}^T\sim4$, see solid lines in Fig.~\ref{fig3}(b) and SI Sec.~\ref{sec:rabi} for a discussion of the other fit parameters, confirming the spin pair is metastable with a spin-dependent lifetime.  
A simple physical interpretation can be obtained if we assume the third state is a pure spin singlet ($S=0$). In this case, spin selectivity naturally arises due to selection rules -- transition from the $ST_0$ ($T_\pm$) state is allowed (forbidden in the absence of spin-orbit coupling), hence $k_{\rm rec}^S>k_{\rm rec}^T$ \cite{Davies1988,Boehme2003}.  

As the inter-pulse delay is increased further, the optical contrast between $ST_0$ and $T_\pm$ will eventually vanish due to their finite lifetimes (related to $k_{\rm rec}^S$ and $k_{\rm rec}^T$) and the additional effect of longitudinal spin relaxation ($T_1$ process).  
To probe this decay, we increase the dark time $\tau$, with and without a final MW pulse, see sequence in Fig.~\ref{fig3}(c). 
The two traces keep increasing but converge together within $\sim10\,\mu$s [Fig.~\ref{fig3}(d), top]. 
By fitting the differential trace [Fig.~\ref{fig3}(d), bottom], we deduce a $1/e$ decay time of $6\,\mu$s, which corresponds to the average spin polarisation lifetime (due to the combined effects of state lifetime and $T_1$) of the emitters contributing to the spin contrast. 

\begin{figure}[tb!]
    \centering
    \includegraphics{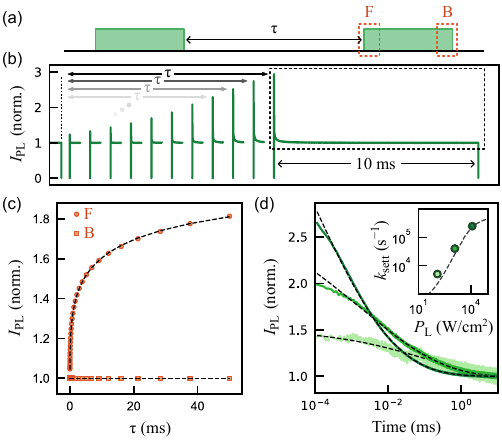}
    \caption{{\bf PL settling-recovery dynamics.} 
    (a)~Pulse sequence consisting of 10-ms laser pulses separated by a variable dark time $\tau$ which allows the system to relax. 
    PL readout is taken from the front ($F$) and back ($B$) of the pulse. 
    (b)~Example PL traces for increasing dark time $\tau$, recorded for an ensemble of emitters. 
    (c)~Integrated PL ($F$, $B$) versus $\tau$. 
    The signal trace $F$ (first 10\,$\mu$s of each laser pulse) is fit with a stretched exponential. 
    The flat reference trace $B$ (last 10\,$\mu$s of each laser pulse) confirms the system is reset to the same state after each pulse independent of $\tau$.  
    (d)~PL traces after $\tau=50$\,ms (ensuring most populations have decayed back to the initial state) plotted on a log time scale, for three laser intensities: $P_{\rm L}=1.2 \times 10^2$, $P_{\rm L}=1.2 \times 10^3$, and $P_{\rm L}=1.2 \times 10^4$ \,W/cm$^2$ (from lighter to darker green). 
    Black dashed lines are stretched exponential fits from which the settling rate $k_{\rm sett}$ is extracted. 
    Inset: Settling rate as a function of $P_{\rm L}$. 
    The dashed line corresponds to the model presented in Fig.~\ref{fig5}(c), with parameters discussed in SI Sec.~\ref{sec:ensemble}.
    }
    \label{fig4}
\end{figure}

As can be seen in Fig.~\ref{fig3}(d), despite all spin polarisation being erased at $\tau=30$\,$\mu$s, the PL is still increasing meaning that some emitters in the ensemble are still trapped in the spin pair configuration. 
To probe the full recovery of all populations into the optically active ground state, i.e.\ the recombination of all the spin pairs in the ensemble, we further increase the dark time $\tau$ preceding the readout laser pulse, see pulse sequence and example PL traces in Fig.~\ref{fig4}(a,b). 
The recovery curve [Fig.~\ref{fig4}(c)] follows a stretched exponential $e^{-(t/T_{\rm rec})^{\beta_{\rm rec}}}$ with a $1/e$ recovery time $T_{\rm rec} \approx 13$\,ms and a stretch exponent $\beta_{\rm rec}\approx0.3$. 
The highly stretched behaviour indicates a range of monoexponential recovery times spanning many orders of magnitude across the defect ensemble under interrogation, from $T_{\rm rec}\sim1$\,s down to $\sim100$\,ns (see SI Sec.~\ref{sec:stretch}). 
Note, in the measurements presented in Figs.~\ref{fig1}-\ref{fig3}, only those emitters with a short spin pair lifetime ($T_{\rm rec}\lesssim 10\,\mu$s) significantly contribute to the PL given the high duty cycle of the laser pulse sequence in these measurements.   

Since the PL recovery was shown to be associated with spin pair recombination, the shelving observed at the start of each laser pulse can logically be assumed to correspond to the inverse process, i.e.\ the creation of the (optically inactive) weakly coupled spin pair. 
To gain insights into the spin pair creation mechanism, the PL settling decay is measured as a function of laser intensity ($P_{\rm L}$), see example PL traces in Fig.~\ref{fig4}(d) plotted on a logarithmic time scale. 
The decay is well fit with a stretched exponential $e^{-(t/T_{\rm sett})^{\beta_{\rm sett}}}$ with $\beta_{\rm sett}\approx0.1$-0.3. 
The settling rate $k_{\rm sett}=1/T_{\rm sett}$ is found to increase roughly linearly with $P_{\rm L}$ [Fig.~\ref{fig4}(d), inset], indicating an excited state population dependent hopping into the optically inactive spin pair configuration.
Like for spin pair recombination, the highly stretched exponential behaviour of the pair creation dynamics suggests the underlying transition rates vary widely across the ensemble.

\begin{figure}[b!]
    \centering
    \includegraphics{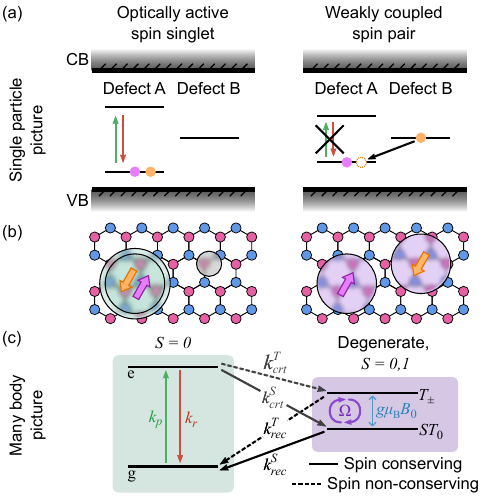}
    \caption{{\bf Proposed electronic structure.} 
    (a)~Single-particle picture for two point defects occupied by two electrons. 
    The green/red arrows represent optical transitions. 
    The black arrow represents a charge transfer from defect B to defect A. 
    (b)~Pictorial representation in the hBN lattice. 
    The two electrons are either co-localised on defect A forming a spin-singlet system (left), or localised on one defect each forming a degenerate singlet-triplet system (right). 
    (c)~Many-body picture of the two-defect system. 
    The four states of the spin pair [see Fig.~\ref{fig1}(g)] are treated as a simplified two-level system (states denoted as $ST_0$ and $T_\pm$) split by $g\mu_{\rm B} B_0$. 
    Spin driving is only enabled between these two states. 
    }
    \label{fig5}
\end{figure}

\section{Defect pair model}

Informed by these observations, we now construct a model of the electronic structure of the system, which we dub the optical-spin defect pair (OSDP) model and is effectively an adaptation of the radical-pair mechanism well established in spin chemistry~\cite{Steiner1989,Woodward2002,
Evans2013}.
We propose the two electrons of the weakly coupled spin pair (once created) are carried by two distinct point defects (potentially residing in distinct hBN layers), defect A and defect B, separated by $\gtrsim1$\,nm.
If the two defects have different electronic structures, they may form a donor-acceptor pair~\cite{Tan2022, Auburger2021} where the electron from e.g.\ defect B will spontaneously hop to defect A [Fig.~\ref{fig5}(a,b)]. 
This hopping corresponds to the recombination of the weakly coupled spin pair to form a closed-shell spin singlet ($S=0$).  
Optical emission is assumed to originate from electronic transitions within defect A when both electrons are co-localised on it, but is disabled in the spin pair configuration which corresponds to a different charge state of defect A. 
The spin pair is created when the excited electron on defect A hops to defect B.
The single-site (defect A) optical transitions, rather than inter-defect donor-acceptor transitions with large separations~\cite{Tan2022, Auburger2021, Dean1973}, support the generally high brightness, narrow emission lines and short lifetime observed for single emitters \cite{GuoNatComms2023, Tran2016, Kumar2023, Pelliciari2024}.

The resulting many-body electronic structure of the OSDP model has four levels in its simplest form, the ground and excited states of the optical cycle (considered to be both pure $S=0$, we ignore the $S=1$ excited state for simplicity), and the two effective spin states of the weakly coupled spin pair ($ST_0$ and $T_\pm$), as depicted in Fig.~\ref{fig5}(c). 
Like for recombination, the spin pair creation process is expected to be spin selective, i.e.\ $k_{\rm crt}^T<k_{\rm crt}^S$.  
The sign of the ODMR contrast depends on the relative ratios of the forbidden/allowed transition rates.
Namely, if the spin pair creation process is less spin selective than the recombination process, i.e.\ $k_{\rm crt}^T/k_{\rm crt}^S>k_{\rm rec}^T/k_{\rm rec}^S$, then $T_\pm$ will get preferentially populated under optical pumping. 
Subsequently, MW-induced spin mixing will lead to an increase in readout PL, hence a positive ODMR contrast. 
Conversely, if $k_{\rm crt}^T/k_{\rm crt}^S<k_{\rm rec}^T/k_{\rm rec}^S$, $ST_0$ will get preferentially populated under optical pumping, and the ODMR contrast will be negative. 
This model can thus account for the variable ODMR sign across emitters observed in previous works~\cite{ChejanovskyNatMat2021, SternNatComms2022, Scholten2023}. 
The model also predicts that despite the two possible signs the ODMR contrast is positive on average, consistent with observations, see further discussion in SI Sec.~\ref{sec:model_odmr}. 

The OSDP model explains the broad range of spin pair recombination [stretched recovery curve, Fig.~\ref{fig4}(c)] and creation rates [stretched settling decay, Fig.~\ref{fig4}(d)], since optical emitters (defect A) across the ensemble will experience a range of distances to their nearest suitable defect B, see further discussion in SI Sec.~\ref{sec:rate_dependence}. 
The distribution of distances between the two defects also explains the large variability in ODMR contrast, from no contrast at all if defect B is far away such that no hopping is possible ($k_{\rm crt}^{S}=0$) to exceeding 100\% for a close-by defect B with high charge hopping rates (e.g.\ $k_{\rm crt}^{S},k_{\rm rec}^{S}\sim10\,\mu$s$^{-1}$, see SI Sec.~\ref{sec:model_odmr}). 
Another interesting prediction of the model is that high ODMR contrasts are associated with short spin pair lifetimes (e.g.\ $1/k_{\rm rec}^{S}\sim100$\,ns), which prevents the observation of Rabi oscillations. 
This explains why Rabi oscillations have only been reported from single emitters with a relatively low ODMR contrast ($\sim1\%$) or from ensembles \cite{GuoNatComms2023, Scholten2023}. 
In SI Sec.~\ref{sec:ensemble}, we determine distributions of the $k_{\rm crt}^{S}$ and $k_{\rm rec}^{S}$ rates which reproduce well all the ensemble-averaged PL data presented in Figs.~\ref{fig2}-\ref{fig4}, as well as the observed average ODMR contrast.

Our OSDP model to explain the origin of ODMR is universal in the sense that it can in principle apply to any optical emitter with an $S=0$ ground state (defect A) that has a suitable nearby donor-like defect (defect B). 
This may explain why ODMR has been observed for hBN emitters with zero-phonon lines ranging from 420\,nm to 800\,nm~\cite{Singh2024}, which likely correspond to defects A with different atomic structures. 
Likewise, due to this universality ODMR-active emitters are expected to be found in a variety of hBN samples, and in other wide bandgap semiconductors.
In SI Sec.~\ref{sec:other_samples}, we investigate three other hBN samples with substantially different physical characteristics to the hBN nanopowder studied above: a hBN micropowder of higher purity sourced from a different supplier, a hBN bulk crystal, and a hBN film grown by metal-organic vapour-phase epitaxy -- all exhibit similar ODMR contrast and photodynamics. 
We also show data obtained from a relatively standard GaN sample (Fig. S4), exhibiting $S=1/2$-like ODMR with a +1\% contrast and very similar photodynamics to hBN, consistent with the OSDP model. 
Additionally, we note that single emitters exhibiting $S=1/2$-like ODMR were recently reported in GeS$_2$ \cite{Liu2024}, which may also be explained by our model.

The OSDP model presented in Fig.~\ref{fig5} can be extended to describe situations where the primary defect (defect A) has a more complex electronic structure. 
In particular, the $S=0$ excited state may experience an intersystem crossing to a metastable $S=1$ state before charge transfer, or if the two electrons occupy degenerate orbitals localised on defect A then the ground state may be a spin triplet ($S=1$). 
Charge transfer to/from a nearby defect B as described by the OSDP model can then provide a mechanism for ODMR of $S=1$ and $S=1/2$-like spin transitions simultaneously, all detected through the same optical emitter. 
This model may thus explain recent reports of these two types of spin resonances being observed on single emitters in hBN \cite{Stern2024,Gao2024} as well as in GaN \cite{Luo2024}.

\begin{figure}[b!]
    \centering
    \includegraphics{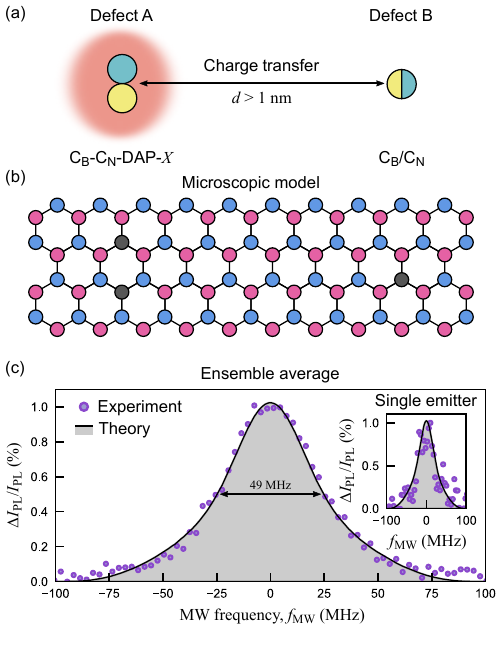}
    \caption{{\bf Microscopic modeling.}
    (a)~A general schematic representation for a proposed example OSDP in hBN.
    The optical emitter, defect A, is a donor-acceptor pair (DAP) consisting of two carbon defects, C$_{\text{B}}$ and C$_{\text{N}}$ separated by a short distance $X$ (normalised by the bond length). 
    A spin pair is formed after charge transfer to defect B, an additional donor or acceptor (either C$_{\text{B}}$ or C$_{\text{N}}$) separated from the DAP by $d>1$\,nm.
    (b)~Lattice representation for a candidate microscopic model consisting of C$_{\text{B}}$-C$_{\text{N}}$-DAP-$2$ and C$_{\text{B}}$ (shown here with $d=1.6$\,nm).
    (c)~Comparison of the experimental ODMR linewidth ($\approx 50$\,MHz) for an ensemble average of many OSDPs and the simulated linewidth ($49$\,MHz) based on the hyperfine interaction. 
    Inset: A single emitter ODMR ($\approx 48$\,MHz linewidth) compared with the same simulation.
    }
    \label{fig6}
\end{figure}

Finally, we discuss the possible microscopic structure of the point defects involved. 
Based on general considerations and first-principles calculations presented in SI Sec. XI, we identify substitutional carbon-based defects as likely candidates for both defects~A and B. 
Carbon is one of the most common contaminants in hBN and is likely present in all our samples~\cite{OnoderaNL2019}. 
Substitutional carbon defects, such as C$_{\text{B}}$ and C$_{\text{N}}$, form optically active donor-acceptor pairs (DAPs) \cite{jara_first-principles_2021,linderalv_vibrational_2021,Auburger2021} that may account for a series of emission lines spanning the visible spectral range \cite{Tan2022,Pelliciari2024}. 
To explain optically addressable spin pairs emitting across a broad spectral region, we propose a model in which a C$_{\text{B}}$-C$_{\text{N}}$ DAP serves as the bright defect (defect~A), coupled to a distinct ($ d > 1$\,nm), single-site C$_{\text{B}}$ or C$_{\text{N}}$ defect (defect~B) [Fig.~\ref{fig6}(a,b)]. 
The positive (negative) charge state of the DAP~+~C$_{\text{B}}$ (DAP~+~C$_{\text{N}}$) defect naturally accommodates a charge-transfer metastable state between the singlet ground and excited states (see SI Sec. XI). 
In this metastable state, both the remote C$_{\text{B}}$ (or C$_{\text{N}}$) defect and the charged C$_{\text{B}}$-C$_{\text{N}}$ DAP possess spin-1/2 states, enabling spin-selective transitions. 
For defect A, we computationally investigate C$_{\text{B}}$-C$_{\text{N}}$-2, C$_{\text{B}}$-C$_{\text{N}}$-$\sqrt{7}$, C$_{\text{B}}$-C$_{\text{N}}$-$\sqrt{13}$, and C$_{\text{B}}$-C$_{\text{N}}$-4 DAPs \cite{Auburger2021}, which emit across a broad spectral range of 500\,nm to 800\,nm. 
Beyond energy level structures and optical properties, the computed 49~MHz electron spin resonance linewidth for the C$_{\text{B}}$-C$_{\text{N}}$-2 + C$_{\text{B}}$ system aligns very well with our ODMR measurements, as shown in Fig.~\ref{fig6}(c). 
Our microscopic model can be generalized to complexes of donors and acceptors incorporating other defects and impurities, such as the donor oxygen substitutional (O$_{\text{N}}$), further enhancing the diversity of optically addressable spin pairs in hBN.
While further theoretical work is necessary to fully analyze these defect pair systems and validate predictions against experimental data, the proposed examples illustrate that the ODMR mechanism can plausibly arise from combinations of ubiquitous hBN defects.

\section{Summary}

In summary, drawing on a logical series of experiments, we uncovered a surprisingly simple and universal model – two nearby point defects with just three electronic states in total – that comprehensively explains all known characteristics of the optically addressable $S=1/2$-like spins in hBN, and why they are so ubiquitous.  
The presented insights into the electronics structure of these systems are a cornerstone for future theoretical and experimental efforts, providing guidance for identifying specific atomic structures and pathways for on demand creation.
Similarly inspired searches in other materials affords potential opportunities for engineering tuneable systems in quantum technology applications.

\section*{Acknowledgements}
This work was supported by the Australian Research Council (ARC) through grants CE200100010, FT200100073, FT220100053, DE200100279, DP220100178, DE230100192, and DP250100973, and by the Office of Naval Research Global (N62909-22-1-2028). 
I.O.R. is supported by an Australian Government Research Training Program Scholarship.
P.R. acknowledges support through an RMIT University Vice-Chancellor’s Research Fellowship. 
V.I.\ acknowledges the support from the National Research, Development and Innovation Office of Hungary (NKFIH) within the Quantum Information National Laboratory of Hungary (Grant No.\ 2022-2.1.1-NL-2022-00004) and Grant No.\ FK145395. 
This project is funded by the European Union under Horizon Europe (projects 101156088 and 101129663). 
First-principles calculations were enabled by resources provided by the National Academic Infrastructure for Supercomputing in Sweden (NAISS) at the Swedish National Infrastructure for Computing (SNIC) at Tetralith, partially funded by the Swedish Research Council through grant agreement No.\ 2022-06725 and KIF\"U high-performance computation units in Hungary.

\section*{Data availability}

The data supporting the findings of this study are available within the paper and its supplementary information files.

\section*{Competing Interests Statement}

The authors declare no competing interests.

\appendix

\section{Experimental details} 

\subsection{Sample preparation} \label{sec:sample}

The hBN sample investigated in this work was hBN nanopowder purchased in 2017 from Graphene Supermarket (BN Ultrafine Powder). 
The as-received powder was subjected to 2\,MeV electron irradiation with a fluence of $10^{18}$\,cm$^{-2}$ which was found to increase the density of optically active defects~\cite{Scholten2023}. 
For the ensemble measurements, the irradiated powder was suspended in isopropyl alcohol (IPA) at a concentration of 20 mg/mL and sonicated for 30 min using a horn sonicator. 
The sediment from the suspension was drawn using a pipette, then drop cast on a printed circuit board (PCB) used for microwave (MW) delivery, forming a relatively continuous film~\cite{RobertsonACSNano2023}.
For the measurement of single emitters, the suspension was further diluted and drop cast on a glass coverslip until single emitters could be easily resolved. The coverslip was then placed on the PCB. 

\subsection{Experimental setup} \label{sec:setup}

The ensemble measurements were carried out on a custom-built wide-field fluorescence microscope. Optical excitation from a continuous-wave (CW) $\lambda = 532$~nm laser (Laser Quantum Opus 2 W) was gated using an acousto-optic modulator (Gooch \& Housego R35085-5) and focused using a widefield lens to the back aperture of the objective lens (Nikon S Plan Fluor ELWD 20x, NA = 0.45). 
The photoluminescence (PL) from the sample was separated from the excitation light with a dichroic mirror and filtered using longpass and shortpass filters, before being sent to either (1) a scientific CMOS camera (Andor Zyla 5.5-W USB3) for imaging and ODMR measurements, or (2) an avalanche photodiode (Thorlabs APD410A) for time-resolved PL measurements, or (3) a spectrometer (Ocean Insight Maya2000-Pro) for PL spectroscopy.

\begin{figure}[tb]
    \centering
    \includegraphics{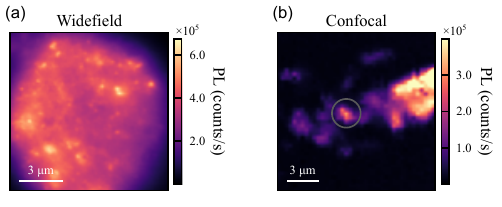}
    \caption{{\bf Example PL images.} 
        (a)~Widefield PL image of the dense powder sample used for ensemble measurements. 
        (b)~Confocal PL map of the dilute powder sample used for measurements of single emitters. 
        Grey circle indicates a single emitter.
        }
    \label{SI_PLmaps}
\end{figure}

The laser intensity at the sample ($P_{\rm L}$) was controlled by the laser power, which had a maximum value of $500$\,mW (going into the objective), with a fixed laser spot size of $100\,\mu$m ($1/e^2$ diameter), see an example PL image in Fig.~\ref{SI_PLmaps}(a).  
A laser intensity of $P_{\rm L}\approx 1.2\times10^4$\,W/cm$^2$, defined as the average power density of a Gaussian beam calculated for the $1/e^2$ beam diameter, was used for the ensemble measurements, except in Fig.~\ref{fig4}(d) where $P_{\rm L}$ was varied. 
For a laser intensity of $P_{\rm L}\sim10^4$\,W/cm$^2$, the typical PL rate at the detector was $\sim10^{10}$ counts/s. 
If we assume a per-emitter PL rate of $\lesssim10^{5}$ counts/s at this laser intensity (conservative estimate based on the counts observed from single emitters), we deduce that the number of addressed emitters in the ensemble is $\gtrsim10^{5}$. 

For the time-resolved PL measurements, the output of the photodiode was digitized with a PicoScope 5244D. 
The bandwidth of the measurement was limited by the photodiode (10\,MHz bandwidth). 
MW excitation was provided by a signal generator (Windfreak SynthNV PRO) gated using an IQ modulator (Texas Instruments TRF37T05EVM) and amplified (Mini-Circuits HPA-50W-63+). 
A pulse pattern generator (SpinCore PulseBlasterESR-PRO 500\,MHz) was used to gate the excitation laser and MW and to synchronise the acquisition. 
The output of the amplifier was connected to the PCB comprising a coplanar waveguide, terminated by a 50\,$\Omega$ termination. 

The external magnetic field was applied using a permanent magnet, and the measurements were performed at room temperature in ambient atmosphere. The only exception is Fig. \ref{fig1}(e,f), for which the sample was placed in a closed-cycle cryostat with a base temperature of 5\,K \cite{Lillie2020} which allowed us to apply a calibrated magnetic field using the enclosed superconducting vector magnet.

Measurement of the single emitters was performed using a confocal microscope equipped with a $\lambda = 532$~nm laser (Laser Quantum Gem 300 mW), and an NA = 0.9 objective (Nikon 100x), see an example PL map in Fig.~\ref{SI_PLmaps}(b). 
Second-order correlation measurements ($g^{(2)}$) were completed using a Hanbury-Brown-Twiss interferometer with avalanche photodiodes (Excelitas SPCM AQRH-14-FC) for single photon detection along with a time correlator (PicoQuant PicoHarp 300). 
PL spectra were measured using Princeton Instruments Acton SP2300 spectrometer. 
For CW ODMR measurements, a signal generator (AnaPico APSIN 4010) and amplifier (Keylink KB0727M47C) provided the MW signal to the PCB. 
An acousto-optic modulator (G\&H AOMO 3080-120) pulsed the laser while a MW switch (Mini-Circuits ZYSWA-2-50DR+) controlled MW delivery for the PL time trace measurements. 
A Swabian Time Tagger 20 was used to time-bin the photon incidences while a SpinCore PulseBlasterESR-PRO 300\,MHz controlled and synchronised each component. 
The estimated laser intensity in the confocal spot was $P_{\rm L}\approx 10^5$\,W/cm$^2$ for these measurements.

For the excited state lifetime measurement presented in the inset of Fig.~\ref{fig1}(d), a separate confocal system was used with excitation from a pulsed laser (NKT Photonics Fianium, wavelength $525\pm30$\,nm, 10\,MHz repetition rate, $5\,\mu$W average power into the confocal spot). 
The PL was detected either with a single-photon counting module (Excelitas SPCM-AQRH-14-FC) and a time tagger (PicoQuant PicoHarp 300) for lifetime measurements, or with a spectrometer (Princeton Instruments IsoPlane). 
30 different PL spots from the sample were measured, showing PL spectra with small differences from spot to spot. 
Averaging the PL spectra from the 30 spots gave a PL spectrum consistent with the ensemble spectrum shown in Fig.~\ref{fig1}(d). 
The fluorescence lifetime was found to vary from spot to spot, ranging from about 1\,ns to 10\,ns, with no apparent correlation with PL intensity or spectral features. 
The data presented in the inset of Fig.~\ref{fig1}(d) is the average of all 30 spots, thus giving the typical lifetime of the defect ensemble. The fluorescence decay was well fit using a stretched exponential function $e^{(-t/\tau_e)^{\beta}}$ where $\tau_e$ is the $1/e$ decay time and $\beta$ is the stretch exponent. 
For the averaged curve, we found $\tau_e=2.0(1)$\,ns and $\beta=0.70(4)$.

\section{Note on the stretched exponential function}\label{sec:stretch}

\begin{figure}[tb]
    \centering
    \includegraphics{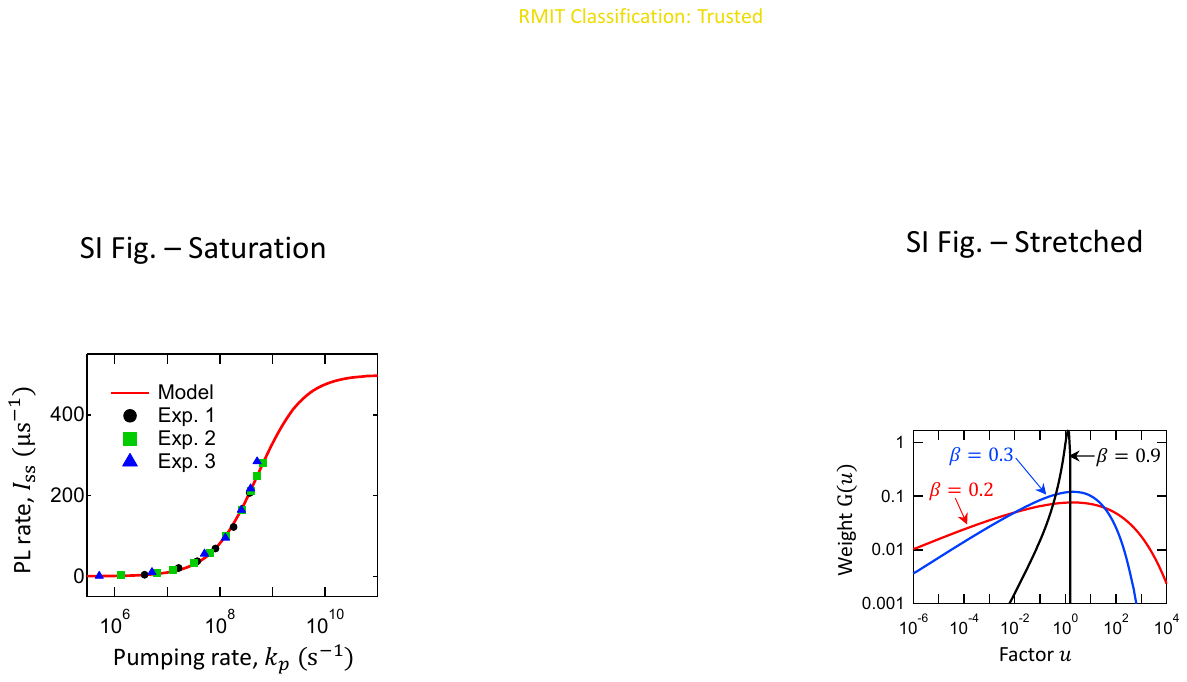}
    \caption{{\bf Stretched exponential function.} 
    Distribution $G(u)$ of the monoexponential components of a stretched exponential function for selected values of the stretch exponent $\beta$.}
    \label{FigSI_stretched}
\end{figure}

In the main text, we found the ensemble-averaged PL traces to be well fit with a stretched exponential $e^{-(t/T_c)^{\beta}}$ where $T_c$ is the $1/e$ decay time and $\beta$ is the stretch exponent.
To understand the meaning of the stretch parameter $\beta$, we recall that a stretched exponential function can be written as a linear superposition of simple exponential decays~\cite{JohnstonPRB2006}, 
\begin{equation} \label{eq:stretched}
e^{-x^\beta}=\int_0^\infty du \rho(u)e^{-x/u}=\int_0^\infty du G(u)\left(\frac{e^{-x/u}}{u}\right)
\end{equation}
where the distribution function $\rho(u)$ is given by
\begin{equation}
\rho(u)=-\frac{1}{\pi u}\sum_{k=0}^\infty \frac{(-1)^k}{k!}\sin(\pi\beta k)\Gamma(\beta k+1)u^{\beta k}\,.
\end{equation}
In Eq.~\ref{eq:stretched}, we introduced another distribution function, $G(u)=u\rho(u)$, such that the basis function $\frac{e^{-x/u}}{u}$ has a unity area under the curve regardless of $u$. 
This is a more meaningful distribution in our case since the area under the curve represents PL counts (when measuring the settling rate), and we want to know the contribution of each component $u$ to the total PL counts. 
The distribution $G(u)$ is plotted in Fig.~\ref{FigSI_stretched} for different values of $\beta$. 
The case $\beta=0.2$, for example, reveals that there are significant contributions (defined as having a weight $G(u)$ exceeding more than half of the maximum weight) for $u$ between $10^{-3}$ (i.e.\ a monoexponential with decay time $10^3$ times shorter than $T_c$) and $10^{+2}$ (monoexponential decay time $10^2$ times longer). 
Such a wide range of rates for the spin pair creation and recombination processes across the ensemble of emitters indicates that the underlying processes vary widely from emitter to emitter. 
This is consistent with our interpretation as a charge transfer process between the optically active defect and a nearby defect at a variable distance, see further discussion in SI Sec.~\ref{sec:rate_dependence}. 

\section{Analysis of other samples and materials} \label{sec:other_samples}

\subsection{Hexagonal boron nitride}

In this paper we focused our analysis on emitters present in hBN nanopowder sourced from Graphene Supermarket (particle size $\sim100$\,nm, purity 99.0\%). 
To confirm our conclusions apply to other hBN samples, we performed a similar investigation on three other hBN samples with substantially different physical characteristics: hBN micropowder sourced from Sky Spring Nanomaterials (particle size 3-4\,$\mu$m, purity 99.99\%), a hBN bulk crystal sourced from HQ Graphene, and a 40-nm-thick hBN film grown by metal-organic vapour-phase epitaxy (MOVPE), see growth details in Ref.~\cite{Chugh2018} and previous optical/spin characterisations in Refs.~\cite{MendelsonNatMat2021, Scholten2023}). 
All three samples were characterised as received or as grown, with no irradiation or annealing performed. 
PL and ODMR from ensembles of visible emitters were observed in all cases, as previously reported~\cite{Scholten2023}. 

\begin{figure*}[tb]
    \centering
    \includegraphics{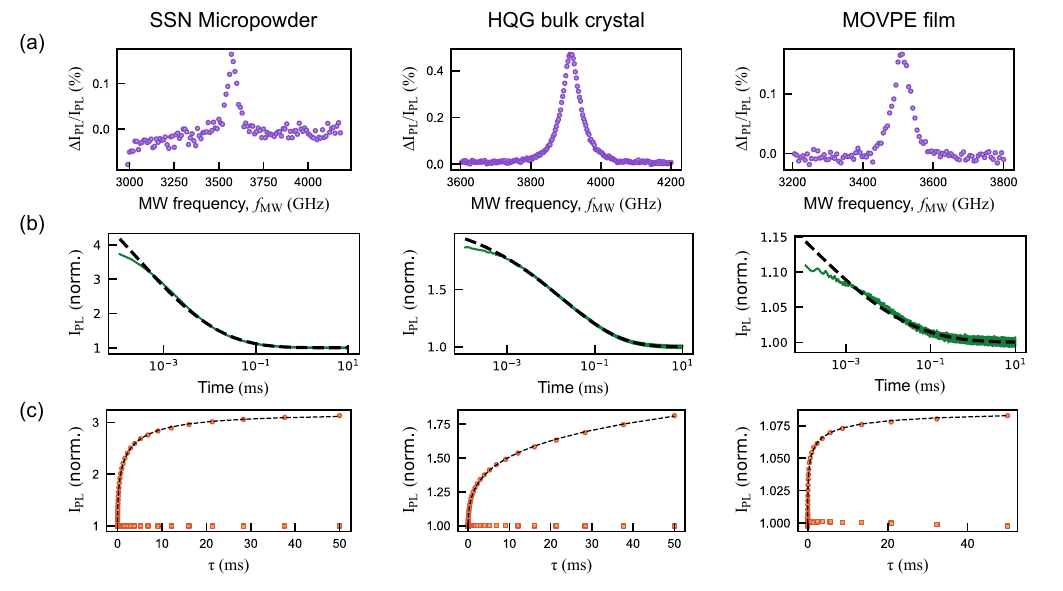}
    \caption{{\bf Analysis of other hBN samples.} 
    (a-c)~Series of measurements performed on ensembles of visible spin defects in three different hBN samples: micropowder sourced from Sky Spring Nanomaterials (SSN, left column), bulk crystal sourced from HQ Graphene (HQG, middle column), and MOVPE-grown film (right column). 
    (a)~ODMR spectrum under a magnetic field $B_0\approx$\,120-140\,mT. 
    (b)~PL trace during a laser pulse after a dark time $\tau=50$\,ms, analogous to Fig.~\ref{fig3}(d). 
    The black dashed line is a stretched exponential fit from which the settling time $T_{\rm sett}$ is extracted. 
    (c)~Integrated PL versus dark time $\tau$, analogous to Fig.~\ref{fig3}(c). 
    Both the signal (rising) and reference (flat) curves are displayed. 
    The signal curve is fit with a stretched exponential (black dashed line) to extract the recovery time $T_{\rm rec}$.}
    \label{SI_other_samples}
\end{figure*}

\begin{table*}[tb]
\centering
\begin{tabular}{M{3 cm} | M{2 cm} | M{2 cm} | M{2 cm} | M{2 cm}}
 \hline
 \hline
 Sample & ODMR contrast (\%) & Overshoot amplitude & $T_{\rm sett}$ ($\mu$s) & $T_{\rm rec}$ (ms)\\ [0.5ex]
 \hline
 GS nanopowder & 0.26 $\pm$ 0.01 & 1.9 & 4 $\pm$ 1  & 13.5 $\pm$ 0.9\\ 
 SSN micropowder & 0.17 $\pm$ 0.02 & 2.7 & 5 $\pm$ 1 & 1.59 $\pm$ 0.03\\
 HQG bulk crystal & 0.475 $\pm$ 0.002 & 0.9 & 39 $\pm$ 1 & 147 $\pm$ 62\\
 MOVPE film & 0.18 $\pm$ 0.01 & 0.1 & 15 $\pm$ 1 & 0.8 $\pm$ 0.1\\
 GaN film & 1.00 $\pm$ 0.01 & 0.2 & 10 $\pm$ 1 & 0.8 $\pm$ 0.1\\
 \hline
 \hline
\end{tabular}
\caption{Summary of parameters extracted from Fig.~\ref{SI_other_samples} for the three additional hBN samples, and from Fig.~\ref{FigSIGaN} for the GaN sample. 
The parameters obtained for the hBN sample analysed in the main text (nanopowder sourced from Graphene Supermarket, GS) are included for comparison. 
The overshoot amplitude (column 3) is the relative amplitude of the PL overshoot peak extracted from Fig.~\ref{SI_other_samples}(b), e.g.\ a value of 0.9 means that the peak amplitude is 90\% of the steady-state PL value.}
\label{table:comparison}
\end{table*}

For each sample, we performed a series of ODMR and time-resolved PL measurements similar to those described in the main text. 
To allow direct comparison between samples, the conditions were kept identical, and the laser intensity fixed to $P_{\rm L}\approx10^4$\,W/cm$^2$. 
The graphs are presented in Fig.~\ref{SI_other_samples} and the extracted parameters summarised in Table~\ref{table:comparison}, where the main text sample was included for comparison. 
The ODMR contrast is relatively consistent across all four samples, lying in the range 0.2-0.5\%. 

Qualitatively similar photodynamics effects are also observed in all samples, namely the PL overshoot and settling as the laser is turned on, and the recovery of these shelved populations in the dark, both of which exhibit highly stretched exponential decays. 
Quantitatively, the samples differ somewhat in terms of the relative amplitude of the PL overshoot and the $1/e$ decay times for the settling and recovery dynamics. 
However, this is expected from our model since each sample will have different types and densities of point defects and so the distribution of the charge transfer rates leading to spin pair creation and recombination will be unique to each sample.

\subsection{Gallium nitride}

\begin{figure}[bt]
    \centering
    \includegraphics{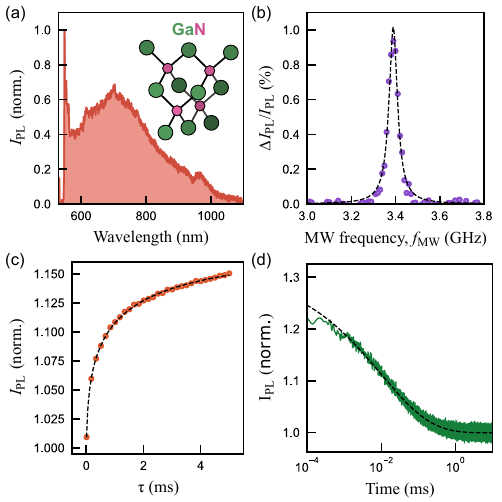}
    \caption{
    {\bf Photodynamic properties of spin defects in gallium nitride.} 
        (a)~PL spectrum for the GaN film. Inset: schematic representation of the gallium nitride (GaN) crystal structure.
        (b)~Ensemble averaged ODMR spectrum of a single $S = 1/2$ resonance in GaN.
        (c)~Integrated PL versus dark time $\tau$, analogous to Fig.~\ref{fig3}(c).
        (d)~PL trace during a laser pulse after a dark time $\tau \approx 5$\,ms, analogous to Fig.~\ref{fig3}(d).  
    }
    \label{FigSIGaN}
\end{figure}

To test the universality of our proposed model we extend our investigation beyond different hBN samples.
Here we consider gallium nitride, a wide bandgap semiconductor (isoelectronic to hBN and other BN compounds) recently demonstrated to host optically addressable spin defects at room temperature~\cite{Luo2024}.
The exact nature of these defects is currently unknown and their spin state has not been conclusively determined as the observed ODMR structures could not be properly reconciled with $S \geq 1$ systems.
Notably, all defects studied in ref.\,\cite{Luo2024} have a prominent $g \approx 2$ resonance with no observable zero-field splitting measurable only under an applied magnetic field, properties which are markedly consistent with the spin-1/2 behaviour of OSDPs in hBN which we have thoroughly considered here.
As in the previous section, we perform a series of measurements on an ensemble of defects in GaN and show their spin and photodynamic properties are consistent with the mechanism described by our OSDP model.

The sample is a GaN film (thickness $\sim1\,\mu$m) grown on sapphire by chemical vapor deposition. 
Under 532\,nm laser excitation, the film emits uniform PL from ensembles of emitters, with a broad spectrum centred around 700\,nm [Fig.~\ref{FigSIGaN}(a)]. 
An ODMR spectrum was measured under an applied field $B_0 \approx 120$\,mT (confirmed with a reference hBN sample, not shown) and a single $g \approx 2$ resonance (linewidth $\approx 50$\,MHz) is observed [Fig.~\ref{FigSIGaN}(b)].
As such we can strongly assert these defects should be associated with an ODMR mechanism compatible with spin-1/2 behaviour.
We next perform time-resolved PL measurements similar to those described in the main text [Fig.~\ref{FigSIGaN}(c,d)] and find qualitatively consistent behaviour with all of the measured samples of hBN.
Specifically, the PL overshoot and settling as the laser is turned on, and recovery of PL from shelved populations while waiting in the dark are observed, with similar stretched exponential behaviour.
It should also be noted the PL overshoot and settling are observed in similar time-resolved measurements for the single defects reported in ref.~\cite{Luo2024}.
Extracted parameters for the ODMR and photodynamics measurements are summarised in Table~\ref{table:comparison} and compare favourably to those of hBN.

\section{Spin dynamics measurements} \label{sec:spin_measurements}

\begin{figure*}[tb]
    \centering
    \includegraphics{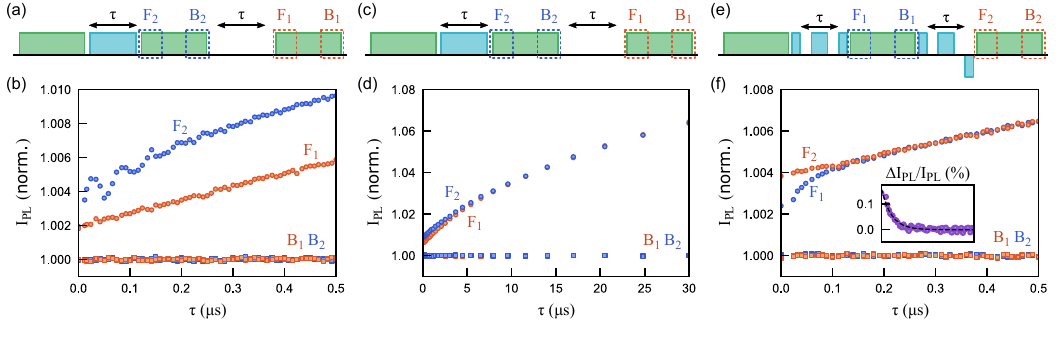}
    \caption{{\bf Spin dynamics measurements.} 
    (a)~Rabi measurement pulse sequence indicating the front ($F_1, F_2$) and back ($B_1, B_2$) gated regions for PL averaging. 
    (b)~$F_1, F_2$ and $B_1, B_2$ plotted against $\tau$. 
    (c)~$T_1$ pulse sequence similarly with $F_1, F_2$ and $B_1, B_2$ marked on the laser pulses which are plotted against $\tau$ in (d). 
    (e)~Hahn echo pulse sequence with $F_1, F_2$ and $B_1, B_2$ marked on the laser pulses which are plotted against $\tau$ in (f). 
    Inset: Normalised Hahn echo data fit with a single exponential.}
    \label{SI_spin_dynamics}
\end{figure*}

Here we show the complete dataset for the spin dynamics measurements presented in Fig.~\ref{fig3}, and additionally show the results of a Hahn echo measurement, see Fig.~\ref{SI_spin_dynamics}.
For each sequence, we gate the recorded PL response and consider only the first ($F_1, F_2$) and last ($B_1, B_2$) $5$\,$\mu$s of the pulse, where the subscripts 1 and 2 refer to two different variants of the pulse sequence: with and without a MW pulse for the Rabi and $T_1$ measurements, and with a different phase of the final $\frac{\pi}{2}$ pulse for the Hahn echo measurement. 
In the main text, we showed only the front-of-the-pulse traces $F_1$ and $F_2$. 
The reference back-of-the-pulse traces $B_1$ and $B_2$ are used to check the laser pulses are sufficiently long to re-initialise the system to the same state (electronic and spin) each time. 

Once this is verified (giving flat traces for $B_1$ and $B_2$), we can confidently use the signal traces $F_1$ and $F_2$ to analyse the effect of the pulse sequence.
To partly remove the recovery component in the signal traces, we can calculate a normalised spin contrast $N=F_2-F_1$.
This provides a better way to estimate the spin dynamics parameters such as the $T_1$ and $T_2$ times. 
However, we note that because the recovery decay is spin dependent ($k_{\rm rec}^S\neq k_{\rm rec}^T$), this normalisation does not completely separate the recovery dynamics from the spin dynamics. 
This is particularly true for the spin lifetime measurement, where the difference $N=F_2-F_1$ decays on a similar time scale to the recovery time of some of the emitters in the ensemble, and therefore the $T_1$ time cannot be simply estimated by monoexponential fitting and instead a comparison to the full photodynamics model is required, see further discussion on this point in SI Sec.~\ref{sec:ensemble}. 
A similar difficulty exists in near-surface nitrogen-vacancy centers in diamond where charge effects can be confused with spin relaxation~\cite{BluvsteinPRL2019}.   

On the other hand, in the Hahn echo measurement the PL recovery effect is much less prominent on the time scale of the spin dephasing [see Fig.~\ref{SI_spin_dynamics}(f)], and so we expect the normalised signal $N=F_2-F_1$ to be representative of spin dephasing only. 
By fitting this normalised signal to a monoexponential decay [Fig.~\ref{SI_spin_dynamics}(f), inset], we infer $T_2 = 54(4)$\,ns.  

\section{Additional data from single emitters}\label{sec:singles}

\begin{figure*}[tb!]
    \centering
    \includegraphics{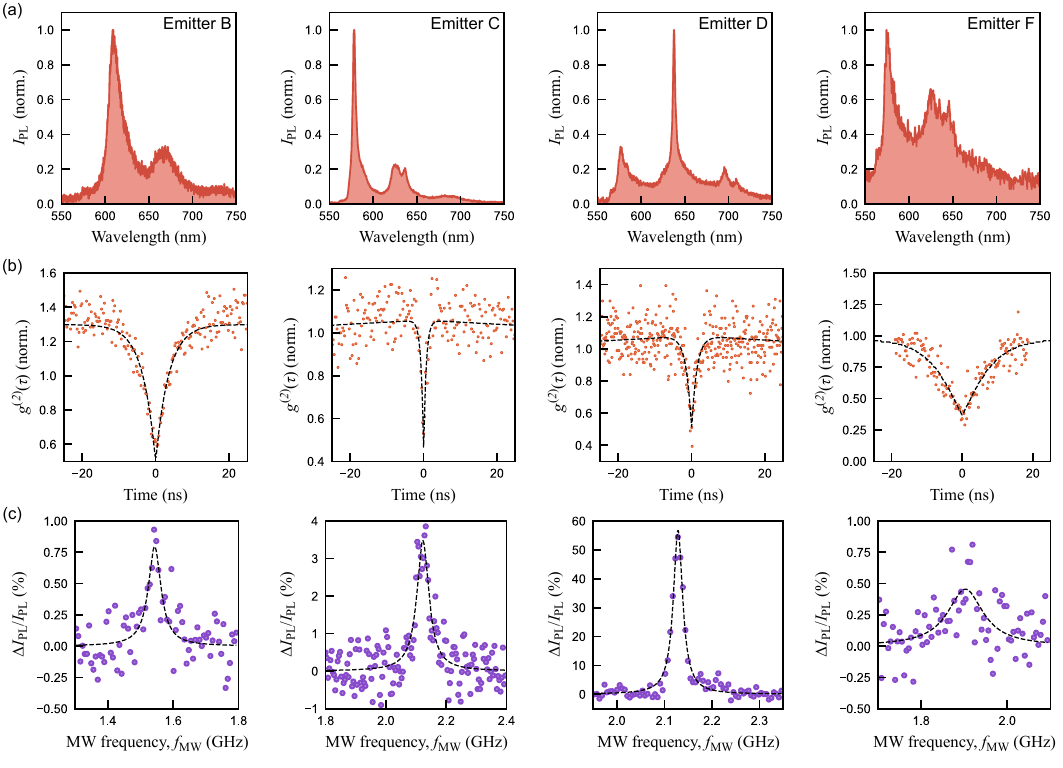}
    \caption{{\bf ODMR-active single emitters.}
    (a)~PL spectra of four selected emitters. 
    (b)~Auto-correlation function of the photon emission for the same emitters as in (a). 
    (c)~ODMR spectrum for the same emitters as in (a) under magnetic fields ranging from $B_0\approx55$\,mT to $B_0\approx75$\,mT.
    }
    \label{SI_singles}
\end{figure*}

\begin{figure}[tb!]
    \centering
    \includegraphics{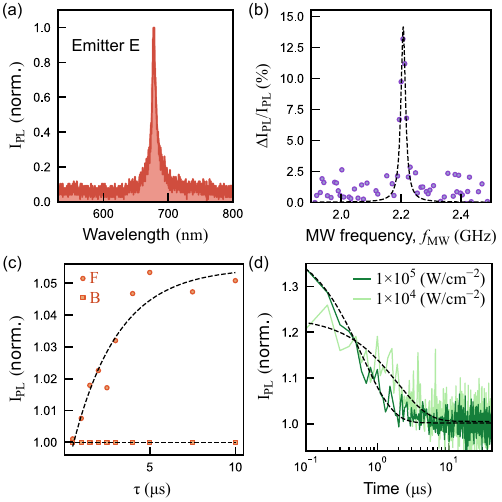}
    \caption{{\bf PL settling-recovery dynamics for a single emitter.} 
    (a)~PL spectrum and (b) ODMR spectrum for emitter E. 
    (c)~PL integrated at the front (rising trace) and back (flat trace) of the pulse, versus dark time $\tau$, analogous to Fig.~\ref{fig4}(c). 
    (d)~PL traces during a laser pulse following a long dark time, for two different laser powers differing by roughly an order of magnitude, analogous to Fig.~\ref{fig4}(d). 
    Dashed lines in (c,d) are monoexponential fits.}
    \label{SI_singles_dynamics}
\end{figure}

\begin{figure}[tb!]
    \centering
    \includegraphics{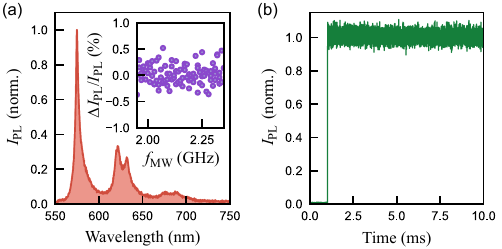}
    \caption{
    {\bf ODMR-inactive single emitter.} 
    ~(a) PL spectrum for a single emitter with no measureable ODMR signature. Inset: ODMR spectrum taken under $B_0 \approx 60$\,mT.
    ~(b) PL trace during a laser pulse after a dark time $\tau = 30$\,ms, analogous to Fig. 3(d).
    }
    \label{SI_single_inactive}
\end{figure}

Here we show additional data from single emitters, measured on the same GS nanopowder as for the ensemble measurements but more diluted to obtain isolated flakes rather than a dense powder film. 
Most emitters did not exhibit measurable ODMR contrast above our noise floor of 0.2\%. However, single emitters with an ODMR contrast ranging from the 0.2\% noise floor up to 60\% were found. 
Figure~\ref{SI_singles} shows the PL spectrum, photon auto-correlation function (to prove the single emitter nature), and ODMR spectrum of four example ODMR-active emitters. 
The ODMR resonance is at the expected frequency $f_r\approx g\mu_{\rm B} B_0/h$, and the linewidth is consistent with the ensemble results. The CW ODMR contrast for these three emitters is 1\%, 3\%, 60\%, and 0.5\% with linewidths between the range $25$ -- $100$\,MHz.
The ODMR spectrum of the 60\% case was also shown in the inset of Fig.~\ref{fig1}(c). 

In Fig.~\ref{SI_singles_dynamics}, we show data for another single emitter for which we also measured the photodynamics in a similar fashion to Fig.~\ref{fig4}. 
The PL and CW ODMR spectra are shown in Fig.~\ref{SI_singles_dynamics}(a,b), featuring a 14\% ODMR contrast.  
The recovery curve [Fig.~\ref{SI_singles_dynamics}(c)] exhibits a monoexponential recovery with a recovery time of $T_{\rm rec}\approx3\,\mu$s.
The short recovery time, which we interpret as a short spin pair lifetime in our model, is consistent with the relatively large ODMR contrast for this emitter, see a discussion of the predicted ODMR contrast in SI Sec.~\ref{sec:model_odmr}. 
Moreover, PL traces at the start of a laser pulse are shown for two different laser intensities in Fig.~\ref{SI_singles_dynamics}(d), revealing a monoexponential decay with a decay rate $k_{\rm sett}$ scaling with laser intensity, as expected from our model.

Finally, in Fig.~\ref{SI_single_inactive} we show data obtained from a representative single emitter which does not exhibit resolvable ODMR. 
Importantly, there is no sign of charge dynamics in the PL traces (no PL overshoot/settling when the laser is turned on, no recovery in the dark). 
This is consistent with our interpretation that ODMR is associated with charge dynamics between the optically active defect and a nearby defect.

\section{Rabi modelling} \label{sec:rabi}

\begin{figure}[ht]
    \centering
    \includegraphics{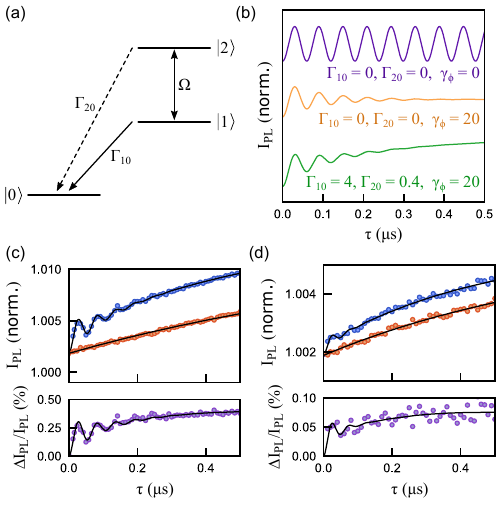}
    \caption{{\bf Rabi modelling and fitting.} 
    (a)~Three level system for modelling the Rabi measurement. 
    (b)~Example Rabi curves from the model with no decay process or dephasing (top curve), with dephasing (middle) and with both decay processes and dephasing (bottom). 
    (c,d)~Fitting of the model to experimental data from the hBN nanopower studied in the main text (c), and  for the MOVPE film (d). 
    The bottom plots are the normalised signal $F_2-F_1$.}
    \label{SI_rabi_model}
\end{figure}

To understand the shape of the Rabi curve we utilise a model in which Rabi oscillations occur between two states with additional decay processes. 
Previous studies have shown openly decaying two-level systems reproduce the observed damped oscillation behaviour \cite{KosugiPRB2005}.
Here we extend upon the two-level system to include a third level to enable PL readout of the spin state. 
Numerical simulations and fitting were facilitated by a Python code adapted from Ref.~\cite{CampaioliPRXQ2024}.

The structure of our model is re-depicted in Fig.~\ref{SI_rabi_model}(a) with states $\ket{0}$, $\ket{1}$, and $\ket{2}$ ($S$, $ST_0$, $T_\pm$ respectively in the main text).
Rates $\Gamma_{10}$ and $\Gamma_{20}$ ($k_{\rm rec}^S$ and $k_{\rm rec}^T$ in the main text) incoherently couple the states $\ket{1}$ and $\ket{2}$ to $\ket{0}$ respectively, and the Rabi frequency $\Omega$ coherently couples states $\ket{1}$ and $\ket{2}$ to each other.
An additional pure dephasing rate, $\gamma_\phi$, is assumed to only be relevant for the states $\ket{1}$ and $\ket{2}$ ($\ket{0}$ has no pure dephasing mechanism) and is not shown.

We express the total Hamiltonian of the system as $H = H_0 + H_1$ where $H_0$ is the energy eigenfunction of the states $\ket{0}$, $\ket{1}$, and $\ket{2}$ which form an orthogonal basis for the system. 
$H_1$ is a perturbation due to an external microwave driving field which coherently couples two of the states ($\ket{1}$ and $\ket{2}$) with Rabi frequency $\Omega$.
In the rotating frame, the Hamiltonian in matrix form is
\begin{align}
    H = \frac{1}{2}\begin{bmatrix}
        0 & 0 & 0 \\
        0 & -\Delta & \Omega \\
        0 & \Omega & \Delta
    \end{bmatrix},
\end{align}
where $\Delta$ is the detuning frequency between the driving field and resonant transition frequency. 
Additionally we have applied the rotating wave approximation which removes fast, out-of-phase oscillations, and taken $\hbar = 1$, used throughout. 

To include the incoherent decay processes to the state $\ket{0}$, we adopt the density matrix formalism and evolve the system using the Linblad master equation, 
\begin{align}
    \Dot{\rho}(t) = -i[H, \rho(t)] + \sum_k \gamma_k \bigg[L_k \rho(t) L_k^{\dagger} - \frac{1}{2} \Big\{ L_k^{\dagger} L_k, \rho(t) \Big\} \bigg],
\end{align}
where $\rho(t)$ is the time dependent density matrix and $\gamma_k$ are transition rates with associated Linblad operators $L_k$.
The Linblad jump operators between the initial state $\ket{i}$ and the final state $\ket{f}$ are defined as $L_{\rm jump} = \ket{f}\bra{i}$ ($i,f = 0, 1, 2$) while pure dephasing operators are defined as $L_{\rm dephasing} = \ket{i}\bra{i}$.

The Linblad equation is solved via the Liouville superoperator, $\mathcal{L}$, and calculating the matrix exponential,
\begin{align}
    \rho(t) = \exp{\big[\mathcal{L}(t - t_0)\big]}\rho(t_0).
\end{align}

To simulate the PL readout in our Rabi measurement, we calculate the intensity as a function of the populations for the three states,
\begin{align}
    I_{\rm PL} = \alpha_0 \rho_{00} + \alpha_1 \rho_{11} + \alpha_2 \rho_{22}
\end{align}
where $\rho_{ii}$ is the population of state $\ket{i}$ taken from the diagonal elements of the density matrix and the coefficients $\alpha_i$ are arbitrary constants relating to the PL intensity of state $\ket{i}$.
Note, in the context of the model presented in Fig.~\ref{fig5}, $\alpha_1=\alpha_2=0$ would be predicted, however, due to the mechanism of optically reading out the spin states with a laser pulse of finite duration, some population cycling occurs during measurement, and thus it is more physically relevant to have $\alpha_0 \gtrsim \alpha_1, \alpha_2>0$.

Three example simulations are shown in Fig.~\ref{SI_rabi_model}(b) for increasing levels of complexity.
Each simulation takes $\rho_{22} = 1$, and $\rho_{00} = \rho_{11} = 0$ as the initial conditions, based on spin selection rules and positive ODMR contrast suggesting $\ket{2}$ is polarised under optical pumping.
In the first example we show the simplest case where there are no decay or dephasing processes ($\Gamma_{10} = \Gamma_{20} = \gamma_\phi = 0$) which recovers the expected oscillatory behaviour for the simple two-level system.
The introduction of dephasing ($\Gamma_{10} = \Gamma_{20} = 0, \gamma_\phi = 20$\,$\mu$s$^{-1}$) causes a damping of the oscillations corresponding to a loss of coherent populations in $\ket{1}$ and $\ket{2}$, but the envelope remains symmetric. 
Finally, we include the incoherent decay processes ($\Gamma_{10} = 4$\,$\mu$s$^{-1}$, $\Gamma_{20} = 0.4$\,$\mu$s$^{-1}$, $\gamma_\phi = 20$\,$\mu$s$^{-1}$) which adds an exponential decay envelope to the curve which qualitatively matches the observed Rabi measurement data.

To better understand the experimental data, we can fit with the numerical model. 
A least-squares minimisation is used to simultaneously fit the PL traces with and without the applied microwave driving for the nanopowder sample [Fig.~\ref{SI_rabi_model}(c), identical to Fig.~\ref{fig3}(b)].
In the fit, all the rates ($\Gamma_{10}, \Gamma_{20}, \gamma_\phi$, and $\Omega$) and PL coefficients ($\alpha_0, \alpha_1$, and $\alpha_2$) are left as free parameters.
The extracted values are $\Gamma_{10} = 3.45$\,$\mu$s$^{-1}$, $\Gamma_{20} = 0.87$\,$\mu$s$^{-1}$, $\gamma_\phi = 24.67$\,$\mu$s$^{-1}$, and $\Omega = 116.26$\,$\mu$s$^{-1}$. 
Additionally we fit another Rabi measurement taken under similar conditions for the MOVPE film discussed in SI Sec.~\ref{sec:other_samples} [Fig.~\ref{SI_rabi_model}(d)] and extract $\Gamma_{10} = 2.89$\,$\mu$s$^{-1}$, $\Gamma_{20} = 1.32$\,$\mu$s$^{-1}$, $\gamma_\phi = 57.56$\,$\mu$s$^{-1}$, and $\Omega = 131.15$\,$\mu$s$^{-1}$.
For both samples the parameters $\alpha_i$ scale to match approximately the normalised PL and follow $\alpha_0 \gtrsim \alpha_1, \alpha_2$ as expected.
The increase in $\Gamma_{10}$ for the MOVPE sample corresponds to a reduction in ODMR contrast consistent with the measured value in Table~\ref{table:comparison}.
Fewer oscillations are observed for the MOVPE sample which may be attributed to increased damping from the higher dephasing rate.

We note the three-level model is an idealised representation of an overall more complex system.
Importantly it does not capture the expected stretched distribution of rates and so it would be inappropriate to assign physical meaning to determined values for $\Gamma_{10}$ and $\Gamma_{20}$. Nevertheless, a general understanding of the system is inferred from the ratio $\Gamma_{20}/\Gamma_{10}$.
In SI Sec.~\ref{sec:ensemble}, we model the Rabi experiment (in the incoherent driving regime for simplicity) using the full four-level model of Fig.~\ref{fig5} including the stretched distributions of rates and taking into account the exact PL readout process, and find good agreement with the data, see Fig.~\ref{FigSIRabiT1}.

\section{Photodynamics modelling} \label{sec:ensemble}

Here we elaborate on the proposed model shown in Fig.~\ref{fig5} and show that appropriate distributions of rates allow us to broadly reproduce all of the ensemble data reported in Figs.~\ref{fig2}-\ref{fig4}. 

\begin{figure}[b]
    \centering
    \includegraphics{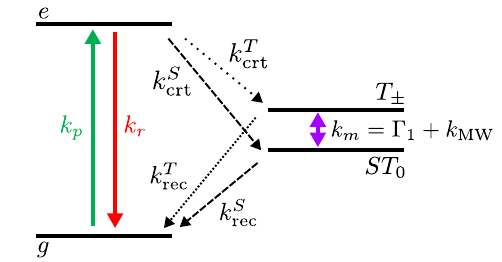}
    \caption{
        \textbf{Photodynamic model of the two-defect system.}
        The system is modelled as a singlet-singlet transition ($g,e$, radiative rates $k_p$ and $k_r$) with creation (crt)/recombination (rec) rates to/from a weakly coupled spin pair configuration ($ST_0, T_\pm$), here simplified as a two-level system.
        The rates coupling the singlet levels to the $T_\pm$ levels ($k^T$) involve a spin flip, and so in general are smaller than the rates coupling to the $ST_0$ states ($k^S$).
        The spin mixing $k_m$ in the spin-pair levels accounts for T$_1$-relaxation ($\Gamma_1$) and MW driving ($k_{\rm MW}$).
    }
    \label{FigSImodel}
\end{figure}

The energy level diagram with annotated rates are reproduced in Fig.~\ref{FigSImodel}. 
Note that an ensemble of defects will have a distribution of these rates. 
However, given the spin pair creation ($k_{\rm crt}^{S,T}$) and recombination ($k_{\rm rec}^{S,T}$) rates are observed to vary significantly ($>5$ orders of magnitude) across the ensemble due to pair displacement [see SI Sec.~\ref{sec:stretch}, \ref{sec:rate_dependence}], we chose to only distribute these parameters to provide rough fits to experimental results, with all other parameters set to singular values. 
We also assume $k_{\rm crt}^{S,T}$ and $k_{\rm rec}^{S,T}$ have correlated distributions, i.e.\ each single emitter will have creation and recombination rates that are both high, or both low, relatively.
We find a log-spaced distribution for $k_{\rm crt}^S$ and $k_{\rm rec}^S$ matches measured behaviour best, with values between $5\times10^4$\,\invs and $2\times10^7$\,\invs, and $3\times10^2$\,\invs and $3\times10^7$\,\invs, respectively. 
The origin of these distributions will be discussed in SI Sec.~\ref{sec:rate_dependence}. 
The spin selectivity ratios $k_{\rm crt}^T / k_{\rm crt}^S$ and $k_{\rm rec}^T / k_{\rm rec}^S$ are kept constant at 11\% and 10\%, respectively, chosen to roughly match the measured CW ODMR contrast for the ensemble.
The two-way spin mixing rate between $ST_0$ and $T_\pm$ is expressed as $k_m = \Gamma_1+k_{\rm MW}=1/2T_1+k_{\rm MW}$ where we assume a $T_1$ time of $T_1=6\,\mu$s and a MW-induced spin mixing rate (where applicable) of $k_{\rm MW}=10\,\mu$s$^{-1}$. 
The radiative rate is taken as $k_r=1/\tau_e=500\,\mu$s$^{-1}$.
The optical pumping rate $k_p$ is related to the experimental laser intensity $P_{\rm L}$ (in W/cm$^2$) via a conversion factor $\eta_{\rm L}=k_p/P_{\rm L}$ which is globally adjusted to best match the data. 
We used $\eta_L=3.4\times10^4$\,s$^{-1}$W$^{-1}$cm$^2$. 
For the laser intensity used in most ensemble measurements, this corresponds to $k_p=400\,\mu$s$^{-1}$.

\begin{figure*}[tb]
    \centering
    \includegraphics[width=0.9\textwidth]{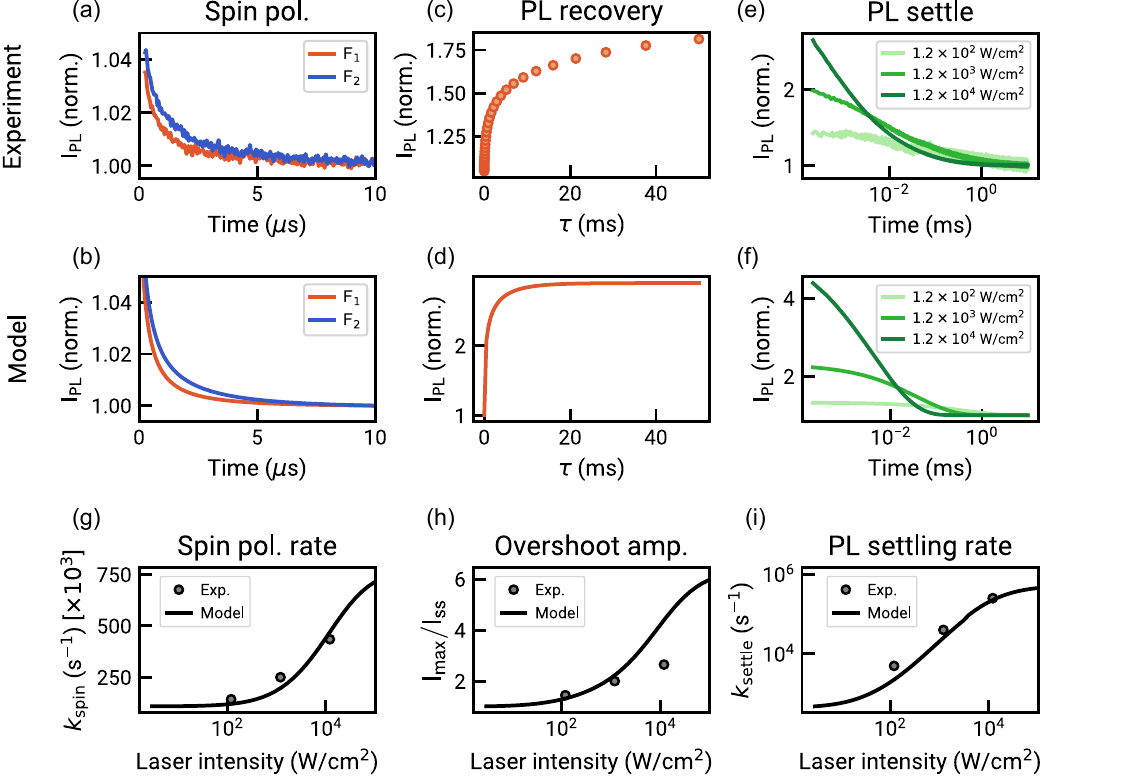}
    \caption{
        \textbf{Comparison of model to ensemble results for the time-dependent PL data.}
        (a,b)~Experimental (a) and simulated (b) PL intensity during a readout laser pulse, with and without a prior MW pulse.
        (c,d)~Experimental (c) and simulated (d) PL at the start of a laser readout pulse, as a function of the dark time $\tau$.
        (e,f)~Experimental (e) and simulated (f) PL during a readout laser pulse, following a long (50\,ms) dark wait duration, for three different laser intensities.
        (g)~Spin polarisation rate fit to curves as in (b), as a function of laser intensity, compared with experimental result.
        (h,i)~Overshoot amplitude (h) and settling rate (i) parameters fit from curves as in (f), as a function of laser intensity, compared with experimental results.
    }
    \label{FigSIModelComp}
\end{figure*}

With these parameters fixed, we compute the PL traces for the various pulse sequences applied, and compare with the experimental data.
We first reproduce the spin-dependent PL traces shown in Fig.~\ref{fig2}(d), corresponding to the pulse sequence Fig.~\ref{fig2}(a). 
Beginning with a relaxed state fully populated in $g$ we apply a long laser pulse to initialise into a (relatively) steady-state population.
After a short dark time ($0.1\,\mu$s) followed by a $1\,\mu$s MW pulse ($F_2$) or additional dark time ($F_1$), a second laser pulse is applied. 
The instantaneous PL emission rate is computed as $I_{\rm PL}(t)=k_rn_e(t)$ where $n_e(t)$ is the excited state population.
The calculated PL traces during the readout pulse are shown in Fig.~\ref{FigSIModelComp}(b), which can be seen to closely match the measured traces [reproduced in Fig.~\ref{FigSIModelComp}(a)]. 
By fitting the difference between the two traces with a monoexponential, the spin polarisation rate $k_{\rm spin}$ can be extracted, which will be discussed later.

To reproduce the PL recovery curve shown in Fig.~\ref{fig4}(c), corresponding to the pulse sequence Fig.~\ref{fig4}(a), we begin with the steady state (following a long laser pulse) and evolve in the dark for a varied time $\tau$, before reading out with a laser pulse.
The PL intensity at the start of the readout pulse (normalised against a short dark evolution) is plot against $\tau$ in Fig.~\ref{FigSIModelComp}(d).
Although the measured recovery shape [Fig.~\ref{FigSIModelComp}(c)] is not reproduced exactly, we see a similar qualitative behaviour with a significant recovery over the first few microseconds and a slower recovery stretching over several milliseconds.
This recovery curve is sensitive to many parameters including $k_r$ and so we expect a better fit would be possible by including distributions of parameters other than just the two considered, beyond the scope of the present study. 

We now reproduce the PL settling traces shown in Fig.~\ref{fig4}(d) by fixing the $\tau$ dark duration to 50\,ms to ensure maximal recovery, and record the PL (in logarithmic time scale) as a function of time during the laser readout pulse  [Fig.~\ref{FigSIModelComp}(f)].
Three values of $k_p$ are modelled, corresponding to the three laser intensities used experimentally [Fig.~\ref{FigSIModelComp}(e)]. 
Again, broad qualitative agreement is found between simulation and experiment.

We now study the dependence on laser intensity of key measurables: the spin polarisation rate $k_{\rm spin}$ extracted from the spin-dependent PL traces [Fig.~\ref{FigSIModelComp}(b)], the amplitude of the initial overshoot ($I_{\rm max}/I_{\rm ss}$, normalised to the steady-state PL) extracted from the PL settling curves [Fig.~\ref{FigSIModelComp}(f)], and the settling rate $k_{\rm sett}$. 
These three quantities are plotted against the pumping rate $k_p$ swept finely, and compared to the values measured for three different laser intensities [Figs.~\ref{FigSIModelComp}(g-i)]
We see broad agreement over many orders of magnitude of the laser intensity, validating the photodynamic model and the parameters listed above.

\begin{figure}[tb]
    \centering
    \includegraphics{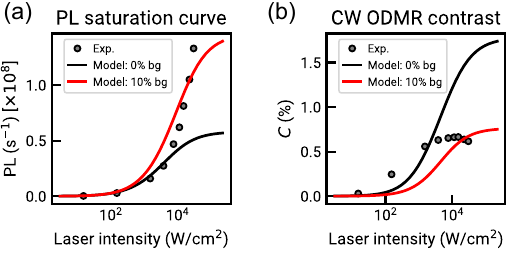}
    \caption{
        \textbf{Modelled laser-dependence of PL and ODMR contrast.}
        (a)~Steady-state PL emission rate as a function of laser intensity, without and with 10\% of additional background emitters (`bg'). All other parameters are the same as in Fig.~\ref{FigSIModelComp}. The $y$-scale of the data is arbitrary.
        (b)~CW ODMR contrast modelled for the same conditions as in (a).
    }
    \label{FigSIcontrast}
\end{figure}

Having matched the model to the time-dependent PL data, we now compare the predictions of the model to the experiment for two steady-state quantities, namely the steady-state PL and the CW ODMR contrast, as a function of laser intensity [Fig.~\ref{FigSIcontrast}].
The steady-state PL is calculated as $I_{\rm ss}=k_rn_e$ where $n_e$ is evaluated as the end of a long laser pulse. 
The CW ODMR contrast is calculated from the steady-state PL with the MW applied ($k_{\rm MW}=10\,\mu$s$^{-1}$), normalised against the steady-state PL with MW off ($k_{\rm MW}=0$),
\begin{align} 
{\cal C}=\frac{I_{\rm ss}(k_{\rm MW}\neq 0)-I_{\rm ss}(k_{\rm MW}=0)}{I_{\rm ss}(k_{\rm MW}=0)}\,.
\end{align}
With the same set of parameters as used for Fig.~\ref{FigSIModelComp}, we see that at high laser intensity the PL and ODMR contrast are under- and overestimated, respectively, compared with experiment.
One possible explanation for this error is that a portion of our measured PL comes from spin-inactive emitters.
To simply model this effect we set 10\% of our emitters to have no decay pathway to the metastable states, i.e.\ $k_{\rm crt}^{S,T} = 0$.
These curves [red in Fig.~\ref{FigSIcontrast}(a,b)] are a closer match to the measured values at high laser intensity, but deviate at low laser intensity. 
We expect that fine tuning this background fraction along with all the other parameters and their respective distributions across the ensemble would lead to a better match, though beyond the scope of this work. 

\begin{figure}[b]
    \centering
    \includegraphics[width=0.98\columnwidth]{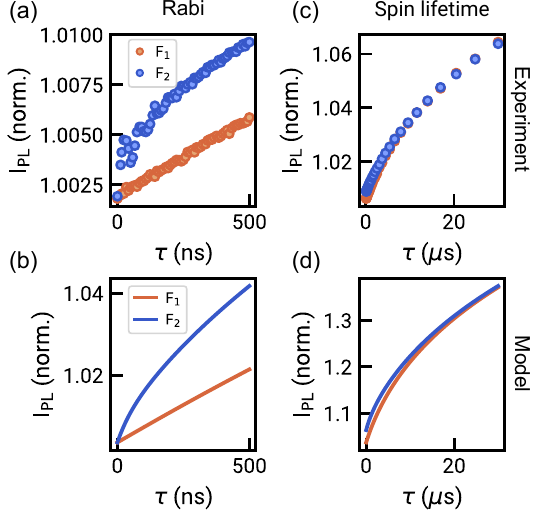}
    \caption{
        \textbf{Modelled Rabi and spin lifetime experiments.}
        (a,b)~Rabi curves with (blue) and without (red) applied MW, measured (a) and modelled (b). Note the model assumes incoherent MW driving and so cannot reproduce the Rabi oscillations.
        (c,d)~Spin contrast decay curves with (blue) and without (red) applied MW, measured (c) and modelled (b).
    }
    \label{FigSIRabiT1}
\end{figure}

As a final test of our photodynamic model, using the same parameters as above (no background emitters), we now simulate the Rabi and spin lifetime experiments as performed in Fig.~\ref{fig3}. 
The modelled Rabi experiment [Fig.~\ref{FigSIRabiT1}(b)] matches the trend of the measured curves [reproduced in Fig.~\ref{FigSIRabiT1}(a)] qualitatively, noting that the model does not include coherence (just a simple mixing rate $k_{\rm MW}=10\,\mu$s$^{-1}$) so cannot reproduce the Rabi flopping oscillations.
Similarly, the modelled spin contrast decay curves match the experimental curves qualitatively [Fig.~\ref{FigSIRabiT1}(c,d)], aside from an overall contrast factor.
The difference between the two traces in Fig.~\ref{FigSIRabiT1}(d) is well fit by a monoexponential with a time constant of 10\,$\mu$s.
Note this simulation assumed an input value for $T_1$ of 6\,$\mu$s, but the apparent value of 10\,$\mu$s indicates a distortion from the spin-dependent recovery as expected. 
Inputting $T_1=3\,\mu$s returns an apparent decay of 6\,$\mu$s as observed experimentally, though this value is sensitive to other parameters preventing a more accurate estimation of $T_1$.

In summary, we found that appropriate correlated distributions of the spin pair creation ($k_{\rm crt}^{S,T}$) and recombination ($k_{\rm rec}^{S,T}$) rates, with all other parameters fixed to singular values, give a good overall agreement with all our experimental data across a range of laser intensities, including the time-dependent PL traces, steady-state PL and ODMR contrast, and the optical readout of the spin dynamics. 
Quantitatively, there are discrepancies between model and experiment in terms of the precise PL amplitude, but given the large parameter space and the many simplifying assumptions, the broad agreement is taken as a strong validation of our model.  

\section{ODMR contrast dependencies} \label{sec:model_odmr}

\begin{figure}[b]
    \centering
    \includegraphics[width=0.85\columnwidth]{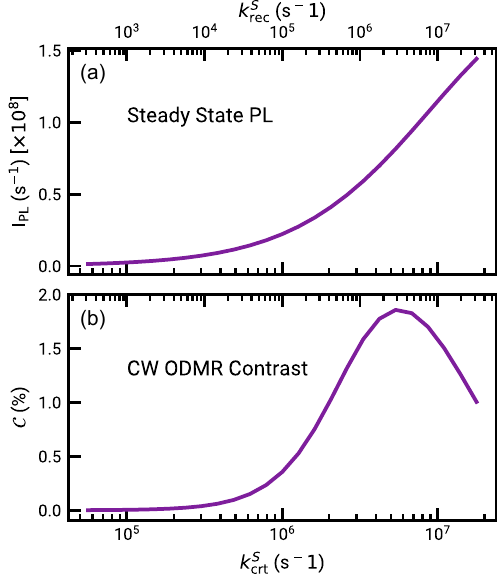}
    \caption{
        \textbf{ODMR contrast of emitters across the ensemble.}
        (a) Steady-state PL and (b) CW ODMR contrast as a function of the correlated pair of rates $k_{\rm crt}^{S}$ (bottom axis) and $k_{\rm rec}^{S}$ (top axis) used to model the ensemble. 
        Other parameters are fixed to: $k_{\rm crt}^T / k_{\rm crt}^S=0.11$, $k_{\rm rec}^T / k_{\rm rec}^S=0.10$, $T_1=6\,\mu$s, $k_{\rm MW}=10\,\mu$s$^{-1}$, $k_r=500\,\mu$s$^{-1}$, $k_p=400\,\mu$s$^{-1}$.  
        }
    \label{FigSIParams}
\end{figure}

An important insight gained from the ensemble modelling performed in the previous section is that the emitters in an ensemble are subject to wide correlated distributions of spin pair creation ($k_{\rm crt}^{S,T}$) and recombination ($k_{\rm rec}^{S,T}$) rates, which relates to the distribution of distances between adjacent point defects. 
While the average ODMR contrast for the ensemble is of order 1\% according to the model, it is interesting to look at the distribution across the emitters composing the ensemble. 
For this, we plot the CW ODMR contrast as a function of the correlated pair of rates $\{k_{\rm crt}^S, k_{\rm rec}^S\}$ used to model the ensemble [Fig.~\ref{FigSIParams}(b)]. 
We see that the contrast is vanishing for small rates and becomes significant ($>0.3\%$) only when $\{k_{\rm crt}^S, k_{\rm rec}^S\}>\{10^6$\,\invs, $10^5$\,\invs\}. 
It reaches nearly 2\% when $\{k_{\rm crt}^S,k_{\rm rec}^S\}\approx\{5\times10^6$\,\invs, $3\times10^6$\,\invs\} before decreasing to 1\% for the highest rates due to the time spent in the spin pair configuration being too short to achieve full spin mixing (given the finite $k_{\rm MW}$). 

From this graph, we infer roughly half of the emitters in the ensemble have an ODMR contrast between 0.3\% and 2\% (fast spin pair cycling), the other half below 0.3\% (slow spin pair cycling). 
However, those emitters with a slow spin pair cycling spend more time trapped in the optically inactive spin state configuration and so contribute less to the detected PL from the ensemble, see dependence of the steady-state PL in Fig.~\ref{FigSIParams}(a), resulting in the 1\% average ODMR contrast calculated for the ensemble.

\begin{figure}[bt]
    \centering
    \includegraphics[width=0.99\columnwidth]{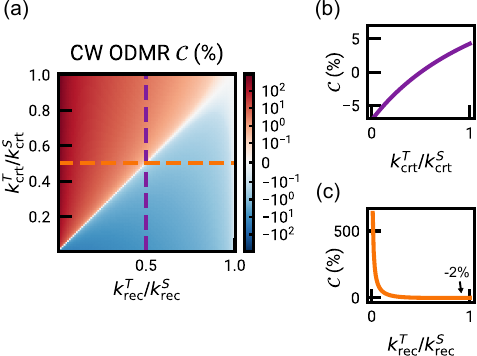}
    \caption{
        {\bf ODMR contrast versus spin selectivity of spin pair creation/recombination.} 
        (a)~Heat map of the CW ODMR contrast as a function of the ratios $k_{\rm crt}^T / k_{\rm crt}^S$ and $k_{\rm rec}^T / k_{\rm rec}^S$, varied from 0.01 to 1.00. Other parameters are fixed to: $k_{\rm crt}^S=8\times10^7$\,s$^{-1}$, $k_{\rm rec}^S=2.5\times10^7$\,s$^{-1}$, $T_1=6\,\mu$s, $k_{\rm MW}=10\,\mu$s$^{-1}$, $k_r=500\,\mu$s$^{-1}$, $k_p=40\,\mu$s$^{-1}$. 
        (b,c)~Linecuts taken along the dashed lines indicated in (a), i.e. along $k_{\rm rec}^T / k_{\rm rec}^S=0.5$ (b) and $k_{\rm crt}^T / k_{\rm crt}^S=0.5$ (c). 
    }
    \label{FigSIHeatmap}
\end{figure}

To explain why some single emitters exhibit a much higher ODMR contrast than displayed in Fig.~\ref{FigSIParams}(b), we need to look at how the contrast varies with other parameters. 
In particular, key parameters governing the contrast are the spin selectivity ratios which were fixed to $k_{\rm crt}^T / k_{\rm crt}^S=0.11$ and $k_{\rm rec}^T / k_{\rm rec}^S=0.10$ in the ensemble modelling. 
In Fig.~\ref{FigSIHeatmap}(a), we plot the ODMR contrast as a function of these ratios as a heat map, with example linecuts shown in Fig.~\ref{FigSIHeatmap}(b,c). 
The contrast is vanishing when the ratios are equal ($k_{\rm crt}^T / k_{\rm crt}^S=k_{\rm rec}^T / k_{\rm rec}^S$), positive when $k_{\rm crt}^T / k_{\rm crt}^S>k_{\rm rec}^T / k_{\rm rec}^S$, and negative when $k_{\rm crt}^T / k_{\rm crt}^S<k_{\rm rec}^T / k_{\rm rec}^S$. 
One can also see that on the positive side the contrast reaches up to 1000\% when $k_{\rm rec}^T / k_{\rm rec}^S=0$, while on the negative side it only reaches down to about -20\% when $k_{\rm crt}^T / k_{\rm crt}^S=0$. 

These behaviours can be understood as follows. 
When the ratios are equal, e.g.\ $k_{\rm crt}^T / k_{\rm crt}^S=k_{\rm rec}^T / k_{\rm rec}^S=0.2$ as an example, the spin pair is polarised in $ST_0$ when created, but because $ST_0$ also decays back to $g$ faster than $T_\pm$ by the same amount, these two effects balance out and on average there is no net spin polarisation and hence no ODMR contrast. 
When $k_{\rm crt}^T / k_{\rm crt}^S<k_{\rm rec}^T / k_{\rm rec}^S$, there is a net polarisation into $ST_0$ under optical pumping. 
MW driving slows down the decay back to $g$ provided $k_{\rm rec}^T / k_{\rm rec}^S<1$, leading to a decreased PL and hence a negative ODMR contrast. 
If $k_{\rm rec}^T / k_{\rm rec}^S=1$, the PL is spin independent and so the contrast vanishes despite spin polarisation. 
When $k_{\rm crt}^T / k_{\rm crt}^S>k_{\rm rec}^T / k_{\rm rec}^S$, there is a net polarisation into $T_\pm$ under optical pumping. MW driving promotes a faster decay back to $g$, hence an increased PL and a positive ODMR contrast.

In summary, within this four-level model, the ODMR contrast of single emitters can reach high positive values such as the 60\% case observed experimentally, provided the spin pair creation and recombination rates are relatively high ($k_{\rm crt}^S, k_{\rm rec}^S\gtrsim 10^7$\,s$^{-1}$) and the spin pair creation process is significantly less spin selective than the recombination process, e.g.\ $k_{\rm crt}^T / k_{\rm crt}^S=0.5$ and $k_{\rm rec}^T / k_{\rm rec}^S=0.1$. 
In addition, we conclude from the heat map Fig.~\ref{FigSIHeatmap}(a) that if these ratios $k_{\rm crt}^T / k_{\rm crt}^S$ and $k_{\rm rec}^T / k_{\rm rec}^S$ are randomly distributed between 0 and 1, then the ODMR contrast is positive on average, e.g.\ taking the average value of the heat map displayed gives +20\%. 
These predictions are consistent with the fact that most hBN samples exhibit positive ensemble contrast, as exemplified in Fig.~\ref{SI_other_samples}. 
As a final comment, we note though the spin selectivity ratios were kept constant when modelling the ensemble data in SI Sec.~\ref{sec:ensemble} as this gave a reasonable agreement with the data overall, it is likely they actually vary from emitter to emitter, with possible correlations with the rates $k_{\rm crt}^S$ and $k_{\rm rec}^S$ hence the distance between the two defects of the pair. 

\begin{figure}[bt]
    \centering
    \includegraphics{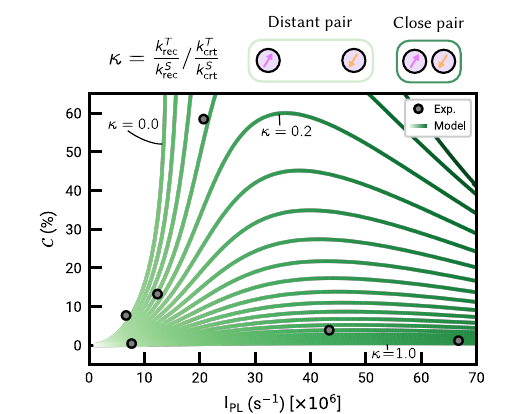}
    \caption{
        {\bf ODMR contrast and PL brightness of single emitters.}
        Scatter plot of ODMR contrast and PL emission rate of measured single emitters (points), compared with predictions from the model (curves). 
        The measured PL count rate was divided by 0.6\% to account for the finite collection efficiency of our experiment and estimate the true emission rate ($I_{\rm PL}$).        
        Each curve was obtained by increasing the rates $\{k_{\rm crt}^S,k_{\rm rec}^S\}$ in a correlated manner as in Fig.~\ref{FigSIParams}. 
        These rates are equivalent to the pair displacement, graphically depicted through the color of the curve. 
        The different curves assume different combined spin selectivity ratios, $\kappa=(k_{\rm rec}^T / k_{\rm rec}^S)/(k_{\rm crt}^T / k_{\rm crt}^S)$. 
        The spin selectivity is varied between $\kappa=0$ (maximally selective), and $\kappa=1$ (no selectivity), from left to right.
        Other parameters are fixed to: $k_{\rm crt}^T / k_{\rm crt}^S=0.5$, $T_1=6\,\mu$s, $k_{\rm MW}=10\,\mu$s$^{-1}$, $k_r=500\,\mu$s$^{-1}$, $k_p=400\,\mu$s$^{-1}$.
    }
    \label{FigSI_modelling_singles}
\end{figure}

To further illustrate that the model is compatible with the data from single emitters, in Fig.~\ref{FigSI_modelling_singles} we compute the PL emission rate $I_{\rm PL}$ and ODMR contrast ${\cal C}$ for a range of model parameters, and compare with the $\{I_{\rm PL},{\cal C}\}$ pairs measured for the 6 single emitters studied in Sec.~\ref{sec:singles}. 
Each curve was obtained by increasing the spin pair creation and recombination rates $\{k_{\rm crt}^S,k_{\rm rec}^S\}$ in a correlated manner as in Fig.~\ref{FigSIParams}, which is physically interpreted as increasing the pair displacement. 
The different curves correspond to different combined spin selectivity ratios, $\kappa=(k_{\rm rec}^T / k_{\rm rec}^S)/(k_{\rm crt}^T / k_{\rm crt}^S)$. 
We find that the experimental results (for PL brightness and ODMR contrast) of single emitters are within the predictions of the model given the two degrees of freedom explored here. 
For instance, the emitter with 60\% ODMR contrast can be explained by relatively fast charge transfer rates $\{k_{\rm crt}^S,k_{\rm rec}^S\}=\{3.0, 0.9\}$\,$\mu$s$^{-1}$ (interpreted as corresponding to a relatively close defect pair) and a high spin selectivity $\kappa\approx0.1$, while emitters with lower ODMR contrast are either remote defect pairs (when they are dim) or close pairs with low spin selectivity $\kappa\sim1$.

\section{On the distance dependence of the transition rates} \label{sec:rate_dependence}

The wide distribution of rates inferred from our measurements may be attributed to different separations between the constituent parts of the OSDP.
Here we briefly suggest the origins for the relationship between the pair separation and transition rates for spin pair creation and recombination.
We consider the scenario where a single optical emitter (defect A) is at the center of its own area of influence with radius $r'$, within which if there is a second proximal defect (defect B), it is able to exchange an electron with it.
The probability of the defect B being within this area is then proportional to $r'^2$.
Given the nature of the weakly coupled spin pair suggests the defects are separated by a distance $\gtrsim 1$\,nm, it is reasonable to assume electron transfer is facilitated by a dipole overlap between the two defects.
Thus, for a defect separation $r$, from Fermi's golden rule, the transition rate is proportional to $1/r^6$.
The large exponent is consistent with a wide distribution of transition rates (e.g.\ the difference between $1$\,nm and $2$\,nm would result in a change by two orders of magnitude).
Additionally, as defect B is more likely to be far from defect A (probability scaling as $r'^2$), the distribution of rates is expected to be weighted towards the slower transition rates. 
This is consistent with the log-spaced distribution for $k_{\rm crt}^S$ and $k_{\rm rec}^S$ used to model the ensemble in Sec.~\ref{sec:ensemble}.

\section{On the effect of spin-spin interactions}

\begin{figure*}[bt!]
    \centering
    \includegraphics{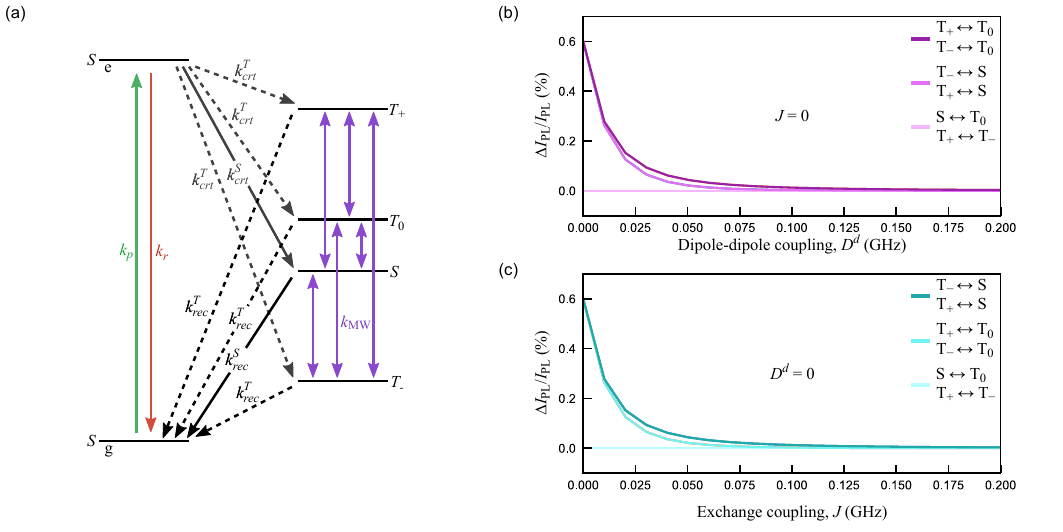}
    \caption{
        {\bf ODMR contrast in the presence of spin-spin interactions.}
        ~(a) Extended energy level diagram for the system with the full four levels of the spin pair ($T_+, T_0, T_-,$ and $S$). 
        The optical singlet states (coupled by radiative rates $k_p$ and $k_r$) are coupled to the spin pair by spin dependent creation ($k^T_{crt}, k^S_{crt}$) and recombination rates ($k^T_{rec}, k^S_{rec}$).
        All levels in the spin pair are coupled to each other by a two-way MW driving rate ($k_{\rm MW}$).
        ~(b) Simulated contrast as a function of the dipole-dipole coupling strength $D^d$.
        ~(c) Simulated contrast as a function of the exchange coupling strength $J$.
    }
    \label{FigSI_JD}
\end{figure*}

So far, for the OSDP system, when modeling the weakly coupled spin pair component we have simplified the system into two levels. 
However, to understand the effects of finite spin-spin interactions on the ODMR contrast, we need to consider all four states of the two-spin system and mix the transition rates following Tetienne et al. \cite{TetienneNJP2012}.
Here we perform simulations of the ODMR contrast as a function of the dipole-dipole and exchange coupling and show that no zero-field splitting for this system is observable as non-negligible coupling otherwise leads to an ODMR-inactive system.

The full system which describes the photodynamic model is depicted in Fig.~\ref{FigSI_JD}(a).
Previous renditions of the model simplified the spin pair into a two-level system whereas here we explicitly consider the four individual states, using the basis of the three triplet states ($T_+, T_0, T_-$) and the singlet state ($S$).
The optical singlet ground and excited states are coupled to each other by radiative rates $k_p$ and $k_r$ and to the spin pair by non-radiative charge transfer rates: $k^S_{crt}$ and $k^T_{crt}$, spin pair creation, and $k^S_{rec}$ and $k^T_{rec}$, spin pair recombination.
As before, the triplet rates $k^T_{crt}$ and $k^T_{rec}$ are considered forbidden and therefore lower probability, thus $k^S_{crt} > k^T_{crt}$ and $k^S_{rec} > k^T_{rec}$.
All states within the spin pair are coupled by a two-way transition rate $k_{\rm MW}$.

CW-ODMR contrast is simulated by solving the eigenvalue problem for the spin pair Hamiltonian,
\begin{multline}
    H = \gamma_e \bold{S}_A \cdot (\bold{B}_0 + \bold{B}_{hyp,A}) + \gamma_e \bold{S}_B \cdot (\bold{B}_0 + \bold{B}_{hyp,B}) \\
    - J \bold{S}_A \cdot \bold{S}_B - D^d[3(\bold{S}_A \cdot \hat{\bold{r}})(\bold{S}_B \cdot \hat{\bold{r}}) - \bold{S}_A \cdot \bold{S}_B],
\end{multline}
where the two spins $A$ and $B$ lie along the vector $\hat{\bold{r}}$, $\gamma_e$ is the gyromagnetic ratio (assumed identical for the two spins), $S_i$ ($i = A, B$) is the spin vector containing the spin operators in the singlet-triplet basis \cite{LimesPRB2013}, $\bold{B}_0$ is the externally applied magnetic field vector, $J$ is the exchange coupling strength, and $D^d$ is the dipole-dipole coupling strength.
We additionally simulate the ODMR linewidth by including a quasi-static hyperfine field by averaging over a zero-mean distribution of fields independent for the two spins, where $\bold{B}_{hyp,i}$ is the hyperfine field vector.
The resultant eigenvector elements are then used to compute the transitions rates in the general case of finite $J$ and $D^d$ as mixtures of the spin-pure creation and recombination rates, similar to how the effect of magnetic fields were modelled in Ref. \cite{TetienneNJP2012}.
However, the state mixing induced by $J$ and $D^d$ also changes the transitions driven by the applied MW field and so to accommodate this we individually scale the MW transition rate $k_{\rm MW}$ for each transition by the corresponding matrix element, inspired by Fermi's golden rule,
\begin{equation}
    k_{\rm MW} \propto \bra{n} \bold{S}_i \cdot \hat{\bold{B}}_1 \ket{m}
\end{equation}
where $n,m = T_+, T_0, S, T_-$ (with $n \neq m$) and $\hat{\bold{B}}_1$ is the MW driving field unit vector.

We simulate the CW-ODMR contrast expected from each of the transitions within the spin pair as functions of both $J$ and $D^d$ (with $D^d=0$ and $J=0$ respectively) as shown in Fig.~\ref{FigSI_JD}(b,c).
For these simulations we have taken the standard deviation for both the hyperfine field distributions as $100$\,MHz (an approximation of the ODMR linewidth), $\hat{\bold{B}}_1 = 1/\sqrt{3}[1, 1, 1]$, and chosen the rates to give a maximum positive contrast of $0.6$\%.
In both cases the transitions $T_0 \leftrightarrow S$ and $T_+ \leftrightarrow T_-$ have zero contrast while all other transitions have maximum contrast for $J = D^d = 0$ which then decays rapidly to zero as the coupling strength is increased.
This is because when $J$ or $D^d$ exceed the linewidth, then the eigenstates are pure singlet and triplet states, with only spin transitions between the triplet states magnetically allowed, and no distinction between these triplet states in terms of the electronic transition rates hence no ODMR contrast. 
As a result, for defect pairs with significant spin-spin interaction (i.e. those separated by $<1$\,nm), even though the splitting between the ESR transitions in principle should be resolvable, there is no longer a mechanism providing spin selectivity and ODMR contrast.

\section{Microscopic model}

In this supplementary section, we perform first-principles calculations to propose a simple microscopic model for optically addressable spin-dependent point defects that aligns with the known features of their spin-dependent decay mechanisms and emitter characteristics. 
Given that spin polarization and spin-dependent decay mechanisms are universal features of OSDPs in hBN and GaN, and our recent work has shown that OSDPs in hBN emit across a broad spectral range, from violet to near-infrared~\cite{Singh2024}, our objective is to identify families of simple defect configurations capable of accounting for multiple emission energies.

It is widely accepted that most emitters in hBN are associated with point defects and their complexes. 
As discussed in the main text, OSDPs are observed in numerous as-received samples from different producers, requiring no additional treatment for activation. 
Consequently, we assume that these emitters are common point defects that form under various growth conditions, remain stable under ambient conditions, and emit across a broad spectral range while maintaining narrow ODMR linewidths.

Recently, donor-acceptor models have been correlated with wide spectral range, high-resolution PL measurements revealing a diverse array of sharp spectral lines~\cite{Tan2022,Pelliciari2024}. Without knowing the microscopic origin of the emitter, the formula for donor-acceptor pairs (DAPs)
\begin{equation} \label{eq:da}
    E_{\text{ZPL}}^{\text{DA}}  = E^{\text{D}} - E^{\text{A}} + \frac{e^2}{4 \pi \epsilon R_{\text{DA}}} \text{,}
\end{equation}
where $E^{\text{D}}$ and $E^{\text{A}}$ are ionization energy levels of the donor and the acceptor, respectively, and $R_{\text{DA}}$ is the distance between the defect pair, could account for many of the PL lines~\cite{Tan2022,Pelliciari2024}.

Donor-acceptor charge transfer may occur during the growth process. Due to Coulomb attraction, charged donors and acceptors tend to pair, potentially leading to the formation of close DA pairs in HPHT growth processes. 
Clustering would cease at this point, as neutral DA pairs do not attract additional donors or acceptors. 
Similarly, donors (or acceptors) already incorporated into the lattice increase the likelihood of incorporating a nearby acceptor (or donor) during epitaxial growth. 
As a result, the formation of the envisioned defect clusters may be independent of the growth conditions, provided that the necessary sources of donor and acceptor defects are present during the growth.

It is well known that  hBN samples are often contaminated with carbon and oxygen in high concentrations, which can reach the percentage scale~\cite{MendelsonNatMat2021}.
Considering only the lowest formation energy single-site carbon and oxygen-related point defects~\cite{weston_native_2018}, we identify C$_{\text{N}}$, C$_{\text{B}}$, and O$_{\text{N}}$ as possible components of optically active DAPs in hBN. 
C$_{\text{B}}$ and O$_{\text{N}}$ act as donors with donor states located at 3.71~eV and 5.33~eV above the valence band edge, respectively, while C$_{\text{N}}$ is an acceptor, exhibiting an acceptor state at 3.19~eV above the valence band edge~\cite{Auburger2021}.

\begin{figure}[h]
  \includegraphics[width=0.9\columnwidth]{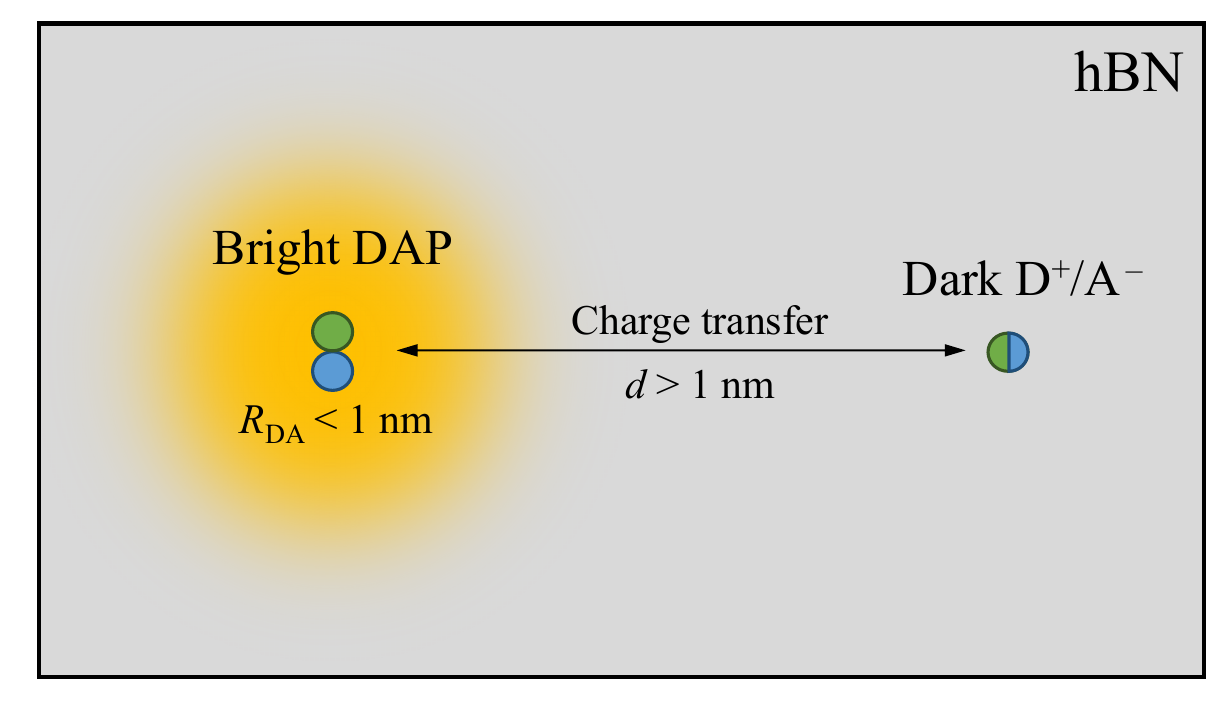}
  \caption{ \textbf{  Schematic representation of the microscopic model for the optical spin defect pair.} The defect system includes a bright, neutral donor-acceptor pair coupled with either a positively charged donor or a negatively charged acceptor.
  } 
  \label{sifigs:dad}
\end{figure}

In our microscopic model, we assume that the optically active defect (defect~A) is a neutral spin-less DAP, see Fig.~\ref{sifigs:dad}. 
Depending on the internal distance of the donor-acceptor pair, such defects can account for a wide range of emitters in the UV-near infrared spectral region~\cite{Tan2022,Pelliciari2024}. 
The internal distance $R_{\text{DA}}$ of the optically active DAP is presumably smaller than 1~nm. To form an OSDP, we include yet another donor (acceptor) of the same kind in its positive (negative) charge state in our microscopic model, see Fig.~\ref{sifigs:dad}. 
The distance $d$ between the bright DAP and the remote dark-charged donor or acceptor is larger than 1~nm to ensure unresolvable splitting of the spin states.

\begin{figure*}[ht]
  \includegraphics[width=0.7\textwidth]{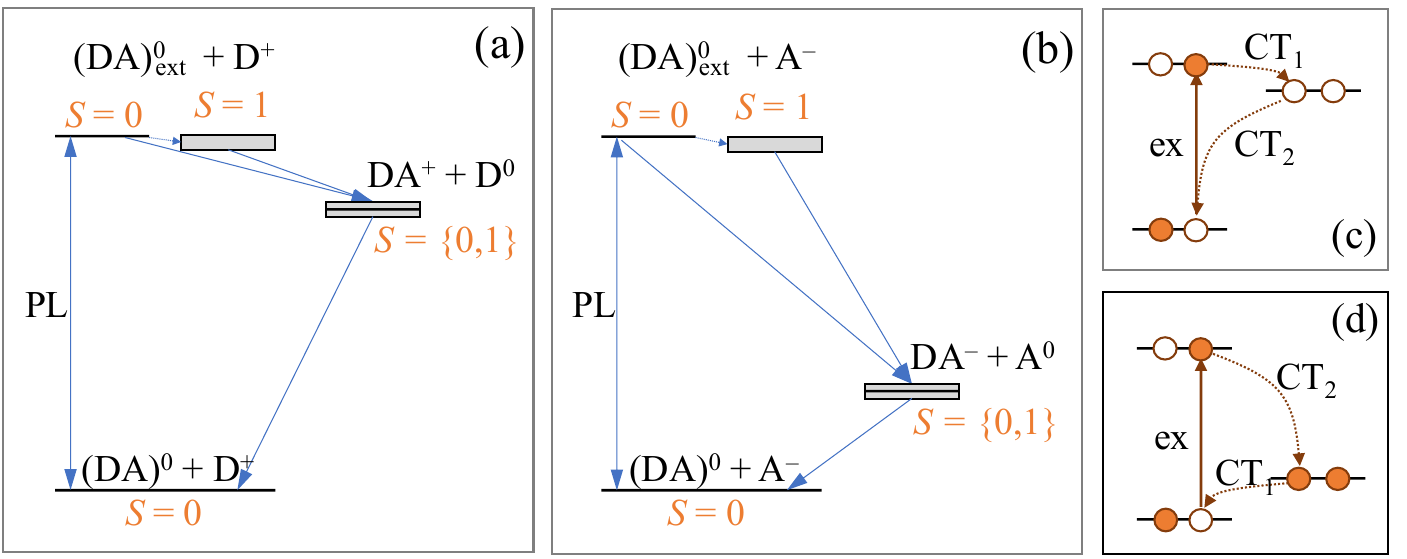}
  \caption{ \textbf{DA$^\text{0}$+D$^\text{+}$ and DA$^\text{0}$+A$^\text{-}$ models for optical spin defect pairs.} 
  Electronic structure and possible transitions of optical spin defect pair, with (a) a remote donors and (b) a remote acceptors. 
  (c) and (d) visualize non-radiative spin selective decay of  the DA$^{0}$+D$^+$ and DA$^{0}$+A$^-$ models. 
  CT$_1$ and CT$_2$ stand for charge transfer processes.
  } 
  \label{sifigs:model}
\end{figure*}

Considering different occupation of the electronic states of the coupled DA$^{0}$+D$^+$  (DA$^{0}$+A$^-$) complex, we can identify three branches of low energy electronic states: the ground state DA$^{0}$+D$^+$ (DA$^{0}$+A$^-$), where DA$^0$ = (D$^+$)-(A$^-$); the optically excited state DA$^{0}_{\text{ext}}$+D$^+$ (DA$^0_{\text{ext}}$+A$^-$), where DA$^0_{\text{ext}}$ = (D$^0$)-(A$^0$); and the charge transferred metastable DA$^+$+D$^0$ (DA$^-$+A$^0$) state, where DA$^+$ = (D$^+$)-(A$^0$) (DA$^-$ = (D$^0$)-(A$^-$)), see Fig.~\ref{sifigs:model}. 
In our model, when $d > R_{\text{DA}}$, the Coulomb interaction ensures, through Eq.~\eqref{eq:da}, that the metastable state lies between the ground and excited states, as the charge transfer to (or from) the distinct donor (or acceptor) leads to a reduction in Coulomb binding energy, see the last term of Eq.~\eqref{eq:da}. 
Assuming that both D$^0$ and A$^0$ have stable spin-1/2 ground state, while the charged D$^+$ and A$^-$ defects are spinless, we find that the ground state is spin-0 and the optically excited and metastable states can give rise to both an open-shell singlet and a triplet spin state. 
Exchange coupling and spin dipole-dipole coupling (zero-field splitting interaction) can cause characteristic splittings between the spin states. 
For close DAPs, both exchange splitting between the singlet and triplet states and zero-field splitting between the triplet spin sublevels are expected, the magnitude of which depends on the DAP distance $R_{\text{DA}}$. 
For the metastable state, we expect negligible exchange and dipolar splitting due to the large $d$ distance between the constituents of the spin pair.

The differences between the DA$^{0}$+D$^+$ and DA$^{0}$+A$^-$ models are illustrated in Fig.~\ref{sifigs:model}. 
The primary distinction lies in the charge transfer mechanism between the bright DAP and the distinct donor or acceptor. 
In the DA$^{0}$+D$^+$ model, an electron may transfer from the excited DAP to the positively charged donor and subsequently decay back to the DAP, as shown in Fig.~\ref{sifigs:model}(c). 
In contrast, for the DA$^{0}$+A$^-$ model, an electron from the negatively charged acceptor may first propagate to the DAP. 
Following this, the excess electron on the DAP decays back to the remote acceptor, as depicted in Fig.~\ref{sifigs:model}(d).

To numerically validate our microscopic model, we performed first-principles calculations on complexes of C$_{\text{N}}$ and C$_{\text{B}}$ point defects using two numerical approaches,  density functional theory (DFT) calculations using supercell models in periodic boundary conditions, and time-dependent density functional theory (TDDFT) on finite hBN flakes.

For the study of periodic models, we use both single-layer models and 3D bulk models consisting of 162 atoms ($9 \times 9 \times 1$ supercell) and  768 atoms ($8 \times 8 \times 6$ supercell), respectively. 
In all cases, we employed experimental lattice parameters, a plane-wave basis set with a cutoff energy of 450~eV, the projector augmented wave (PAW) method~\cite{blochl_projector_1994}, the HSE06 functional~\cite{heyd_hybrid_2003} with a 0.32 mixing parameter~\cite{weston_native_2018}, and the DFT-D3 method with the Becke-Johnson damping function for dispersive corrections~\cite{grimme_effect_2011}. 
The calculations were performed using the Vienna Ab initio Simulation Package (VASP)~\cite{kresse_efficiency_1996}. 
For excited states, we constrained the electronic occupations and calculated the zero-phonon line (ZPL) energies using the $\Delta$SCF method~\cite{jones_density_1989}. 
Hyperfine values were obtained from bulk supercell calculations, including a core polarization correction~\cite{szasz_hyperfine_2013}.

For the study of the flake models, we apply linear response time-dependent DFT  method~\cite{runge1984density,Casida,Dreuw2005} using the ORCA 5.0.3 program suite~\cite{neese2022software}. 
The modeling is carried out using def2-TZVP computational basis set~\cite{Weigend2005} and  PBE0 hybrid functional~\cite{PerdewMix} considering dispersion corrections~\cite{grimme_effect_2011}.
The absorption oscillator strength of the states is computed from the corresponding transitional electric dipole moment. 
The cluster geometries, tailored from a large pristine hBN lattice, are terminated by hydrogens whose bond distance from the neighboring boron and nitrogen atoms is set to the standard value of 1.19 and 1.01~$\AA$ respectively.
During the geometry relaxation, the position of the hydrogens and the attached atoms were kept fixed in order to mimic the compression effect of the embedding hBN lattice on the constructed cluster model~\cite{benedek2024}.

To balance accuracy and computational efficiency in the production calculations, we employed a single-layer cluster model for the distinct DAP systems. 
We keep at least one extra ring of boron-nitride surrounding the minimal model of the DAP system and avoid any potential frustration of hydrogens, i.e., neighboring boron and nitrogen atoms are not allowed to be replaced by the capping hydrogens. 
In the case of the DA$^{0}$+D$^+$ and DA$^{0}$+A$^-$ models, where defect centers are separated by a large distance, we keep the parallelogram shape of the hBN supercell and the geometry is optimized on the cheaper  def2-SVP basis set. 
We tested the convergence of our cluster model with respect to  geometry size and dimensionality, as well as basis set size finding that the above-described minimal model with def2-TZVP basis set gives a reasonable description of the excitations.  
Our wave-function-based analysis confirms that the singlet ground state is of single-reference character and the low-lying excitations are one-electron excitations that can be reasonably treated by the TDDFT theory using the PBE0 functional.

\begin{table}[h]
\setlength\extrarowheight{2pt}
\centering
\caption{ Zero-phonon line energies for four C$_{\text{B}}$C$_{\text{N}}$  DAPs calculated using different methods. 
The table includes HSE(0.32) results reported in Ref.~\cite{Auburger2021}, our HSE(0.32) results obtained using the $\Delta$SCF method with an energy correction for the open-shell singlet excited state, and TDDFT results calculated with the PBE0 functional. }
\label{sitab:ZPL} 
\begin{tabular}{l|ccc}
\hline \hline
DAPs & $\Delta$SCF  & $\Delta$SCF+corr. & TDDFT  \\
 & Ref.~\cite{Auburger2021}  & this work & this work 
\\ \hline
DAP-2 &  2.43  & 2.70  & 2.98\\
DAP-$\sqrt{7}$ &  1.98 & 2.08 & 2.10\\ 
DAP-$\sqrt{13}$ &  1.66  & 1,66  & 1.76\\ 
DAP-$4$ &  1.55  &  1.58 & 1.47 \\ 
\hline \hline
\end{tabular} 
\end{table}

First, we examine the properties of the optically active DAPs (defect A). C$_{\text{B}}$C$_{\text{N}}$ DAPs have been studied theoretically in the literature~\cite{jara_first-principles_2021, linderalv_vibrational_2021, Auburger2021}, providing the foundation for our analysis. 
As shown by Auburger et al.~\cite{Auburger2021}, neutral C$_{\text{B}}$C$_{\text{N}}$ DAPs can emit light across a broad spectral range, from the UV to the near-IR. 
Table~\ref{sitab:ZPL} presents the calculated ZPL energy values for C$_{\text{B}}$C$_{\text{N}}$ DAPs considered in our study. 
Depending on the method employed, we obtain slightly different excitation energies, highlighting the uncertainties in DFT-calculated ZPL energies for hBN. 
Compared to the results in Ref.~\cite{Auburger2021}, our calculations yield consistently larger ZPL values, despite using the same functional. 
This difference can be attributed to the inclusion of an energy correction for the open-shell singlet excited state, which was omitted in Ref.~\cite{Auburger2021}. 
The TDDFT results further increase the calculated ZPL values, due to the use of the PBE0 functional with long-range exact exchange coupling. 
The computed values are sensitive to the choice of functional. 
For a more accurate description, functional tuning is required, as demonstrated in the case of the blue emitter in hBN~\cite{ganyecz_first-principles_2024}, although this is beyond the scope of the current first-principles study. 
We expect the true ZPL energies to lie in between our $\Delta$SCF+corr.\ and TDDFT values.

\begin{table}[h]
\setlength\extrarowheight{2pt}
\centering
\caption{Characteristic of the optical excited state of the C$_{\text{B}}$C$_{\text{N}}$ DAPs, listing TDDFT PBE0 ZPL energies, exchange constant $J$ defining half of the energy splitting between the singlet and triplet DAP excitations, and oscillator strength $f$. }
\label{sitab:osc} 
\begin{tabular}{lc|ccc}
\hline \hline
DAPs & $R_{DA}$ ($\rm \AA$) & ZPL (eV) & $J$ (eV) & $f$ \\ \hline
DAP-1 & 1.4 & 4.70 &  0.63 & 0.49 \\
DAP-2 & 2.9 &  2.98 &  0.31  & 0.15\\
DAP-$\sqrt{7}$ & 3.8 &  2.10  & 0.12 & 0.06\\ 
DAP-$\sqrt{13}$ & 5.2 &  1.76  & 0.10 & 0.08 \\
DAP-4 & 5.8 &  1.47  & 0.08 & 0.08 \\
DAP-5 & 7.3 & 1.20 &  0.05 & 0.06  \\
DAP-7 & 10.2 &  0.48 &  0.02 & 0.02  \\
DAP-8 & 11.6 &  0.45 &  0.01 & 0.01  \\ 
\hline \hline
\end{tabular} 
\end{table}

\begin{figure}[h]
  \includegraphics[width=0.8\columnwidth]{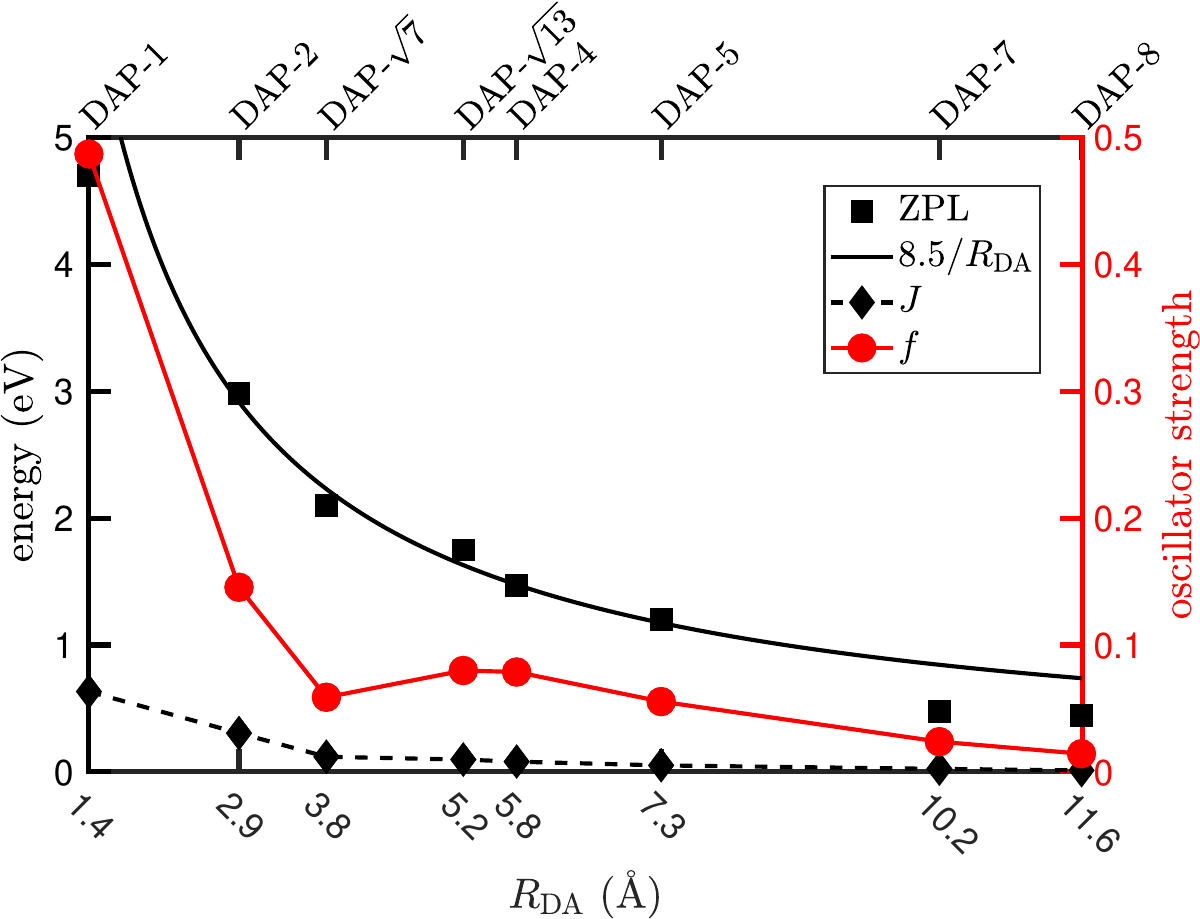}
  \caption{\textbf{Characteristics of DAP excitations as a function of the C$_{\text{B}}$-C$_{\text{N}}$
  distance ($\mathbf{R_{\text DA}}$).} The left y-axis shows the singlet zero-phonon line (ZPL) energy and the exchange constant ($J$), defined as half the energy difference between the singlet and triplet defect electronic excitations. 
  The $8.5/R_{\rm AB}$ function is plotted to verify Eq.~\ref{eq:da} of the ZPL profile. The right y-axis represents the singlet oscillator strength.
  } 
  \label{sifigs:DAP_vs_dist}
\end{figure}

The optical properties of the C$_{\text{B}}$C$_{\text{N}}$ DAPs are further analyzed in Table~\ref{sitab:osc} and Fig.~\ref{sifigs:DAP_vs_dist} using the TDDFT (PBE0) method. 
By examining the dependence of the DAP ZPL energies on $R_{\text{DA}}$ over a wider range, we identify a close $R_{\rm DA}^{-1}$ dependence for intermediate $R_{\rm DA}$ distances, in agreement with Eq.~\eqref{eq:da}. 
We attribute the deviation from the $R_{\rm DA}^{-1}$ fit at short distances to the breakdown of the point charge approximation, which is assumed in the derivation of Eq.~\eqref{eq:da}. 
For large DAP separations (i.e., $R_{\rm DA} > 10~{\rm \AA}$), we find that the employed TDDFT approach tends to underestimate the ZPL energies due to the emergence of the charge-transfer character of the transition, as previously reported in the literature~\cite{maitra_2017}.

We further analyze the excited-state properties of the C$_{\text{B}}$C$_{\text{N}}$ DAPs by calculating the exchange energy $J$ and the absorption oscillator strength $f$, as shown in Table~\ref{sitab:osc} and Fig.~\ref{sifigs:DAP_vs_dist}. $J > 0$ indicates an energetically favorable triplet state in our convention. 
The exchange energy determines the singlet-triplet splitting ($2J$) in the excited state, as depicted in Fig.~\ref{sifigs:model}. 
As observed, the exchange energy decreases rapidly with the $R_{\text{DA}}$ distance between the C$_{\text{B}}$ donor and C$_{\text{N}}$ acceptor. 
At around 1~nm, it drops below our numerical accuracy and becomes undeterminable. Regarding the oscillator strength, we observe a significant reduction from DAP-1 to DAP-2. 
However, in the visible-to-near-IR range covered by DAP-2, DAP-$\sqrt{7}$, DAP-$\sqrt{13}$, and DAP-4, the variation in oscillator strength is relatively minor. 
This suggests that C$_{\text{B}}$C$_{\text{N}}$ DAPs in the visible-to-near-IR region may still be bright enough to remain observable. 
The non-monotonic decay in oscillator strength is expected to result from the orientation dependence of the transition dipole moment between the overall D$_{\text{3h}}$ symmetric orbitals of C$_{\text{B}}$ and C$_{\text{N}}$.

\begin{table*}[ht]
\setlength\extrarowheight{2pt}
\centering
\caption{
Excitation spectrum of the DA$^{0}$+D$^+$ and DA$^{0}$+A$^-$ microscopic models based on C$_{\text{B}}$ donor (D) and C$_{\text{N}}$ acceptor (A). 
The columns list the dominant electronic transitions involved in the excitation process, the weight of the electronic excitation in the excited state, the spin of the excited states, the simulated vertical excitation energies (electronic energy differences without structural relaxation), and the corresponding oscillator strengths. 
The single-particle electronic states contributing to these transitions are illustrated in Fig.~\ref{sifigs:DAA-DAD-MO}. 
The spectrum obtained from our large TDDFT simulations shows good agreement with the electronic structure outlined in Fig.~\ref{sifigs:model}.
}
\label{tab:DAP2_distant_DA} 
\begin{tabular}{c|cccccc|c|c}
\hline \hline
\multirow{2}{*}{model} & \multicolumn{5}{c}{vertical excitation properties}\\
 & dominant excitation  & weight(\%) & spin & energy (eV) & $f$\\
\hline

\multirow{4}{*}{DAP-2 + C$_{\text{B}}^+$} &  \multirow{2}{*}{$b_2$(DAP-2)  $\rightarrow~$$a_2''$(C$_{\text{B}}$)} &\multirow{2}{*}{ 100} &  0 & 1.96 & 0.00 \\
  & & & 1 & 1.96 & - \\
  \cline{2-6}
 &   \multirow{2}{*}{$b_2$(DAP-2)  $\rightarrow~$$b_2^*$(DAP-2)} &\multirow{2}{*}{ 95} & 0 & 3.43 &0.24 \\
  & & & 1 & 2.85 & - \\
  
\hline
\multirow{4}{*}{DAP-2 + C$_{\text{N}}^-$} &  \multirow{2}{*}{$a_2''$(C$_{\text{B}}$)  $\rightarrow~$$b_2^*$(DAP-2)} &\multirow{2}{*}{ 100} &  0 & 2.22 & 0.00 \\
  & & & 1 & 2.22 & - \\
   \cline{2-6}
 &   \multirow{2}{*}{$b_2$(DAP-2)  $\rightarrow~$$b_2^*$(DAP-2)} &\multirow{2}{*}{ 98} & 0 & 3.16 &0.21 \\
  & & & 1 & 2.54 & - \\
\hline \hline
\end{tabular}
\end{table*}

\begin{figure*}[ht]
  \includegraphics[width=0.99\textwidth]{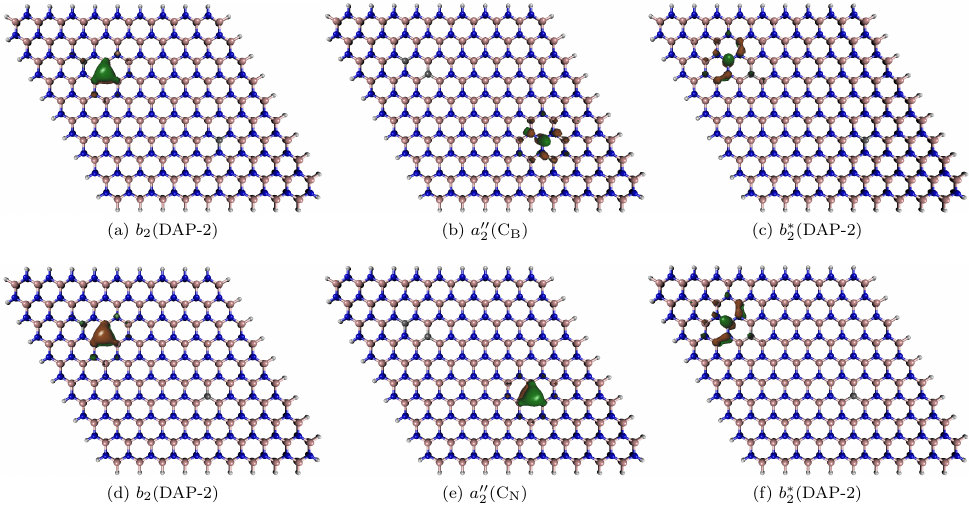}
  \caption{\textbf{Isosurface of defect orbitals.} Plots (a,b,c) and (d,f,g) show the defect orbitals of C$_{\text{B}}$C$_{\text{N}}$-DAP-2  + C$_{\text{B}}^+$  and DAP-2 + C$_{\text{N}}^-$  systems, respectively. 
  Carbon, boron, nitrogen, and terminating hydrogen atoms are colored in gray, pink, blue, and white, respectively.
  } 
  \label{sifigs:DAA-DAD-MO}
\end{figure*}

Next, we investigate possible realizations of the DA$^{0}$+D$^+$ and DA$^{0}$+A$^-$ complexes in a single first-principles model, considering C$_{\text{B}}$C$_{\text{N}}$-DAP-2 as defect A and either a single C$_{\text{B}}^+$ or C$_{\text{N}}^-$ defect as defect B positioned $\sim$1.4~nm from the DAP-2. 
In our analysis, we focus on the low-lying vertical excitation spectrum obtained using the TDDFT method (PBE0), as summarized in Table~\ref{tab:DAP2_distant_DA}. 
The defect states are primarily formed by the p$_z$ orbitals of the carbon atoms, with varying contributions from the p$_z$ orbitals of neighboring boron and nitrogen atoms, as shown in Fig.~\ref{sifigs:DAA-DAD-MO}. 
Table~\ref{tab:DAP2_distant_DA} provides details of the low-energy transitions observed in our simulations, including transition weight, excited-state spin, vertical energy, and oscillator strength.

Our numerical results confirm the theoretical energy level structure sketched in Fig.~\ref{sifigs:model}. 
Specifically, 1) the lowest-lying excitation corresponds to the dark charge-transfer state between DAP-2 and the distant donor/acceptor site, with energy that is spin-independent; 2) the second excitation is the optically active single-electron excitation among the defect orbitals of DAP-2.
Furthermore, we find that the vertical excitation energies for the singlet and triplet states align with those of the corresponding standalone DAP-2 model, indicating that the remote donor/acceptor has a negligible effect on the transition energy of the DAP-2 defect. 
Interestingly, the oscillator strengths of local defect excitations in the DA$^{0}$+D$^+$ and DA$^{0}$+A$^-$ complexes are slightly larger than those of the stand-alone DAP-2 defect. 
These observations provide strong support for our general microscopic model and align with experimental findings.

\begin{figure}[ht]
  \includegraphics[width=1.0\columnwidth]{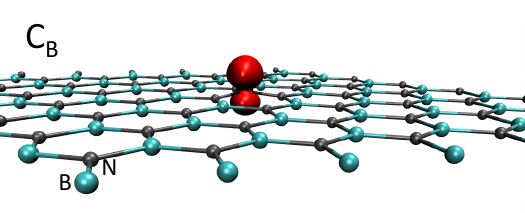}
  \caption{ \textbf{Spin density of an isolated neutral C$_{\text{B}}$ defect in hBN.} The electronic state located in the band gap of hBN is predominantly formed by the p$_z$
  orbital of the substitutional carbon atom.
  } 
  \label{sifigs:spind}
\end{figure}

Last but not least, we examine the spin properties of the DA$^{0}$+D$^+$ and DA$^{0}$+A$^-$ complexes. 
Since the spin density of single substitutional carbon atoms is predominantly localized on the carbon p$_z$ orbital, with secondary localization appearing on the first-neighbor nitrogen and boron atoms (see Fig.~\ref{sifigs:spind}), we expect narrow ESR/ODMR lines for $^{12}$C-related DA$^{0}$+D$^+$ and DA$^{0}$+A$^-$ complexes. 
Meanwhile, $^{13}$C-including complexes may give rise to characteristic splittings that could help identify the defects \cite{Gao2024}.

To confirm this hypothesis, we employ large-scale periodic supercell models to minimize finite-size effects \cite{takacs_accurate_2024} and calculate the hyperfine tensors for all relevant nuclear spins around the neutral C$_{\text{B}}$ and C$_{\text{N}}$ defects, i.e., defect B in the metastable state, and the positively charged C$_{\text{B}}$C$_{\text{N}}$-DAP-2, representing an example of defect A in the metastable state. 
Hyperfine tensors with hyperfine splittings $A_z$ greater than 1~MHz are provided in Tables~\ref{tab:hypCB}, \ref{tab:hypCN}, and \ref{tab:hypDAP2}, respectively.

Considering C$_{\text{B}}$, we obtain a sizable 274~MHz hyperfine splitting. 
It is important to note that the carbon atom in the C$_{\text{B}}$ configuration tends to relax outward, which changes the character of the state from p$_z$ toward an sp hybridization. 
This relaxation significantly influences the hyperfine splitting value. In a single-layer model, where the carbon atom can relax outward without the confining potential of neighboring layers, we obtain an even higher hyperfine splitting of $A_z = 342$~MHz. 
We expect that the hyperfine splitting of the $^{13}$C$_{\text{B}}$ defect may vary over a broad range depending on local strain and temperature. 
The spin density is slightly less localized on the carbon atom in the case of the C$_{\text{N}}$ defect, resulting in a smaller hyperfine splitting of $A_z = 155$~MHz for $^{13}$C. 
We note that C$_{\text{N}}$ tends to remain in-plane, suggesting a more stable hyperfine splitting parameter for this defect.

\begin{figure}[h]
  \includegraphics[width=0.8\columnwidth]{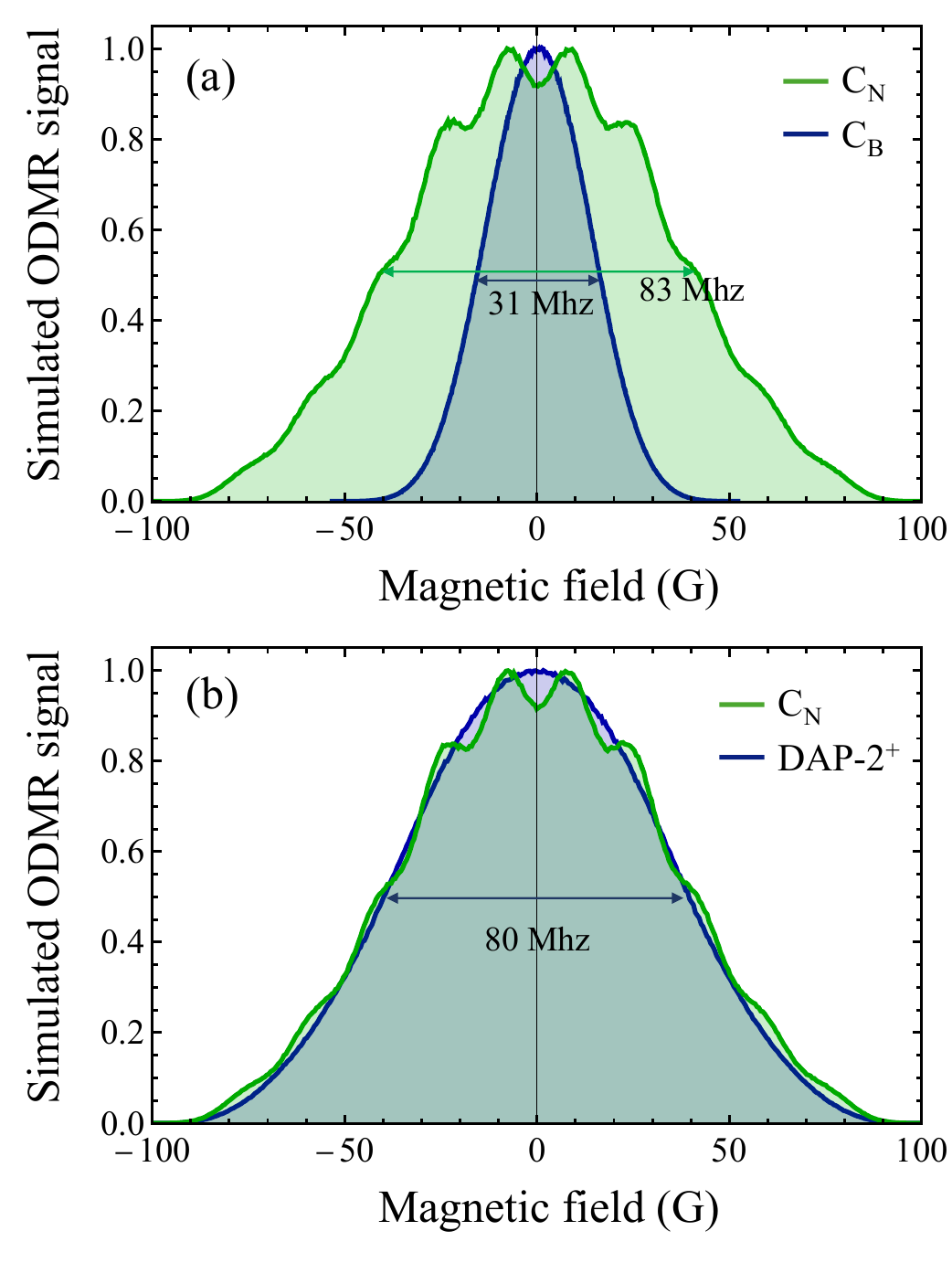}
  \caption{ \textbf{Simulated line profiles for C$_{\text{B}}$, C$_{\text{N}}$, and C$_{\text{B}}$C$_{\text{N}}$-DAP-2$^+$.} Panels (a) and (b) compare the spectra of the neutral C$_{\text{B}}$ and C$_{\text{N}}$ defects, and the C$_{\text{B}}$C$_{\text{N}}$-DAP-2$^+$ and C$_{\text{N}}$ defect, respectively. 
  The theoretical ODMR lines were obtained by taking solely the hyperfine interaction into account. $A_z > 1$~MHz for all included nuclear spins.  
  } 
  \label{sifigs:esr-1}
\end{figure}

As shown in Table~\ref{tab:hypCB}, the $^{11}$B and $^{14}$N nuclear spins couple weakly to the electron spin of the C$_{\text{B}}$ defect. 
As a result, we predict a narrow, 31~MHz wide ODMR spectrum in the absence of carbon nuclear spin for this defect, see Fig.~\ref{sifigs:esr-1}. 
Due to the slight delocalization of spin density for the C$_{\text{N}}$ defect, we observe enhanced hyperfine splitting for the first-neighbor $^{11}$B nuclear spins, resulting in a wider 80~MHz ODMR peak, see Fig.~\ref{sifigs:esr-1}. 
The D$_{3h}$ symmetry of the defect also leads to a resolvable fine structure for the C$_{\text{N}}$ defect. 
In the case of C$_{\text{B}}$C$_{\text{N}}$-DAP-2$^+$, the spin density is primarily located on the C$_{\text{N}}$ site, perturbed by the nearby C$_{\text{B}}$ defect. 
The reduced symmetry lifts the equivalence of symmetrically equivalent sites, see Table~\ref{tab:hypDAP2}, resulting in the loss of hyperfine structure in the simulated ODMR spectra, see Fig.~\ref{sifigs:esr-1}.

\begin{figure}[h]
  \includegraphics[width=0.8\columnwidth]{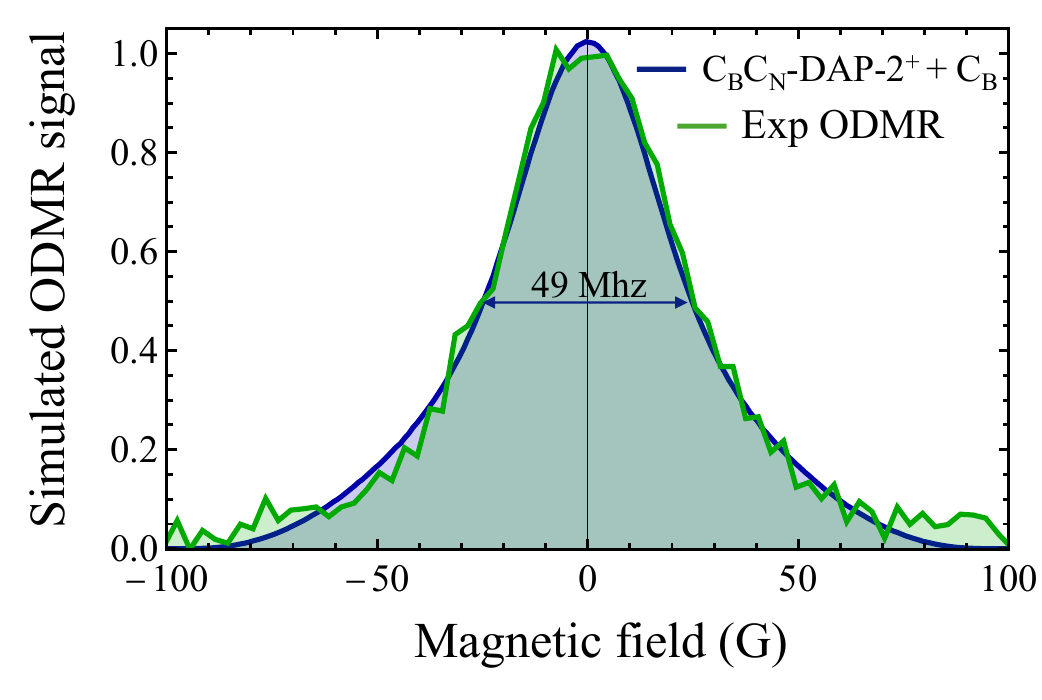}
  \caption{ \textbf{Comparison of the simulated ODMR line profile of C$_{\text{B}}$C$_{\text{N}}$-DAP-2$^+$~+~C$_{\text{B}}$ with sample ODMR data.} The theoretical data have not been fitted, except for normalization and shifting the peak center. 
  } 
  \label{sifigs:esr-2}
\end{figure}

To conclude our discussion on the spin properties, we compare a sample experimental ODMR line (taken from the MOVPE film in Fig.~\ref{SI_other_samples}, whose linewidth is representative of most hBN samples we studied) with the simulated ODMR line obtained by combining, with equal weights, the ODMR lines of the C$_{\text{B}}$C$_{\text{N}}$-DAP-2$^+$ defect and the C$_{\text{B}}$ defect, in accordance with the general model of OSDPs. 
As shown in Fig.~\ref{sifigs:esr-2}, we achieve a perfect match by applying only trivial adjustments, such as normalization and shifting the peak center. 
This observation supports our microscopic model for OSDPs.

Finally, we note that we could easily extend our microscopic model to include additional types of donors and acceptors, such as C$_2$C$_{\text{B}}$, C$_2$C$_{\text{N}}$, and O$_{\text{N}}$. 
In each case, we expect an additional series of ZPL lines from bright DAPs that may couple with distinct donors and acceptors to form optically active OSDPs. 
The oxygen donor, which exhibits a high donor level, is particularly noteworthy, as the C$_{\text{N}}$O$_{\text{N}}$ DAP could emit at shorter wavelengths. 
These will be studied in future work.

\begin{table*}[h]
\setlength\extrarowheight{2pt}
\centering
\caption{Hyperfine coupling tensors of the C$_{\text{B}}$ defect. The results are obtained on HSE(0.32) level of theory using 768 atom bulk hBN supercell. 
The values are obtained for $^{11}$B and $^{14}$N isotopes.}
\label{tab:hypCB} 
\begin{tabular}{c|cccccc|c}
\hline
Atom &  $A_{xx}$ &  $A_{yy}$ &  $A_{zz}$ &  $A_{xy}$ &  $A_{xz}$ &  $A_{yz}$ &  $A_z$ \\ \hline
C & 54.746 & 54.747 & 274.417 & 0.002 & -0.053 & -0.041 & 274.417 \\ \hline
B & 1.114 & -1.114 & 4.664 & -0.385 & -0.110 & -0.210 & 4.670 \\
B & -0.891 & 0.890 & 4.663 & 0.772 & -0.237 & 0.010 & 4.669 \\
B & 1.115 & -1.112 & 4.662 & 0.384 & 0.110 & -0.208 & 4.668 \\
B & -0.223 & 0.224 & 4.661 & 1.157 & -0.126 & 0.199 & 4.667 \\
B & -0.892 & 0.888 & 4.660 & -0.772 & 0.231 & 0.014 & 4.665 \\
B & -0.225 & 0.223 & 4.659 & -1.157 & 0.127 & 0.193 & 4.664 \\
B & -0.296 & -0.475 & 0.861 & 0.155 & 0.525 & 0.303 & 1.053 \\
B & -0.565 & -0.207 & 0.859 & -0.000 & 0.000 & -0.607 & 1.052 \\
B & -0.296 & -0.475 & 0.859 & -0.155 & -0.525 & 0.303 & 1.051 \\
B & 0.516 & 0.107 & 1.018 & -0.355 & 0.004 & -0.001 & 1.018 \\
B & -0.098 & 0.721 & 1.018 & -0.000 & 0.001 & 0.004 & 1.018 \\
B & 0.511 & 0.101 & 1.014 & 0.355 & -0.003 & -0.002 & 1.014 \\ \hline
N & -6.485 & -6.473 & -1.472 & -0.010 & -0.750 & -0.433 & 1.708 \\
N & -6.481 & -6.468 & -1.469 & 0.011 & 0.748 & -0.433 & 1.704 \\
N & -6.460 & -6.487 & -1.468 & -0.000 & -0.001 & 0.864 & 1.703 \\
N & -0.351 & -0.529 & 1.468 & 0.154 & -0.016 & -0.010 & 1.468 \\
N & -0.619 & -0.263 & 1.467 & -0.000 & -0.002 & 0.018 & 1.467 \\
N & -0.352 & -0.530 & 1.466 & -0.154 & 0.015 & -0.011 & 1.466 \\
N & 0.573 & 0.573 & 1.281 & 0.000 & -0.000 & 0.000 & 1.281 \\
\hline
\end{tabular} 
\end{table*}

\begin{table*}[h]
\setlength\extrarowheight{2pt}
\centering
\caption{Hyperfine coupling tensors of the C$_{\text{N}}$ defect. The results are obtained on HSE(0.32) level of theory using 768 atom bulk hBN supercell. 
The values are obtained for $^{11}$B and $^{14}$N isotopes}
\label{tab:hypCN} 
\begin{tabular}{c|cccccc|c}
\hline
Atom &  $A_{xx}$ &  $A_{yy}$ &  $A_{zz}$ &  $A_{xy}$ &  $A_{xz}$ &  $A_{yz}$ &  $A_z$ \\ 
\hline
C & -17.251 & -17.249 & 154.681 & 0.001 & -0.025 & -0.041 & 154.681 \\
\hline
B & -19.517 & -22.564 & -16.456 & 2.640 & 0.026 & 0.015 & 16.456 \\
B & -19.517 & -22.565 & -16.453 & -2.640 & -0.025 & 0.013 & 16.453 \\
B & -24.075 & -17.983 & -16.448 & 0.000 & -0.000 & -0.025 & 16.448 \\
B & -0.733 & -1.666 & -1.768 & 0.308 & -0.003 & -0.002 & 1.768 \\
B & -0.732 & -1.665 & -1.767 & -0.308 & 0.003 & -0.002 & 1.767 \\
B & -1.164 & -1.230 & -1.766 & -0.558 & 0.003 & -0.002 & 1.766 \\
B & -1.164 & -1.230 & -1.766 & 0.558 & -0.003 & -0.002 & 1.766 \\
B & -1.696 & -0.697 & -1.765 & -0.250 & 0.000 & 0.003 & 1.765 \\
B & -1.695 & -0.696 & -1.764 & 0.250 & -0.000 & 0.003 & 1.764 \\
B & -0.245 & -0.244 & 1.503 & -0.000 & 0.000 & -0.002 & 1.503 \\
B & -0.260 & -0.259 & 1.467 & -0.000 & 0.000 & 0.002 & 1.467 \\
\hline
N & -0.226 & -0.820 & 2.780 & -0.003 & 0.006 & 0.002 & 2.780 \\
N & -0.226 & -0.820 & 2.777 & 0.003 & -0.006 & 0.002 & 2.777 \\
N & -0.673 & -0.371 & 2.772 & -0.255 & -0.005 & 0.004 & 2.772 \\
N & -0.674 & -0.372 & 2.770 & 0.255 & 0.005 & 0.004 & 2.770 \\
N & -0.669 & -0.377 & 2.768 & 0.259 & -0.000 & -0.004 & 2.768 \\
N & -0.669 & -0.377 & 2.768 & -0.258 & 0.000 & -0.005 & 2.768 \\
N & 0.168 & 0.287 & 1.102 & 0.000 & -0.000 & 0.316 & 1.146 \\
N & 0.250 & 0.191 & 1.086 & -0.051 & 0.272 & -0.157 & 1.131 \\
N & 0.250 & 0.191 & 1.086 & 0.051 & -0.271 & -0.157 & 1.130 \\
N & 0.151 & 0.263 & 1.041 & -0.000 & 0.000 & -0.299 & 1.083 \\
N & 0.234 & 0.178 & 1.034 & -0.048 & -0.258 & 0.149 & 1.076 \\
N & 0.232 & 0.176 & 1.029 & 0.048 & 0.257 & 0.148 & 1.071 \\
\hline
\end{tabular} 
\end{table*}

\begin{table*}[h]
\setlength\extrarowheight{2pt}
\centering
\caption{Hyperfine coupling tensors of the positively charged C$_{\text{B}}$C$_{\text{N}}$-DAP-2 defect. 
The results are obtained on HSE(0.32) level of theory using 768 atom bulk hBN supercell. 
The values are obtained for $^{11}$B and $^{14}$N isotopes}
\label{tab:hypDAP2} 
\begin{tabular}{c|cccccc|c}
\hline
Atom &  $A_{xx}$ &  $A_{yy}$ &  $A_{zz}$ &  $A_{xy}$ &  $A_{xz}$ &  $A_{yz}$ &  $A_z$ \\ 
\hline
C & -17.177 & -17.230 & 149.670 & 0.047 & -0.044 & -0.037 & 149.670 \\
C & -0.153 & -0.913 & 3.331 & 0.659 & 0.002 & 0.001 & 3.331 \\
\hline
B & -18.367 & -21.170 & -17.648 & 2.427 & 0.003 & 0.002 & 17.648 \\
B & -23.477 & -17.278 & -14.072 & -0.329 & -0.006 & -0.004 & 14.073 \\
B & -19.114 & -21.643 & -14.068 & -2.849 & -0.007 & -0.004 & 14.068 \\
B & -2.305 & -1.001 & -2.035 & -0.231 & -0.000 & 0.002 & 2.035 \\
B & -1.527 & -1.779 & -2.035 & -0.680 & 0.001 & -0.001 & 2.035 \\
B & -0.768 & -1.599 & -1.860 & 0.294 & -0.000 & -0.000 & 1.860 \\
B & -1.136 & -1.231 & -1.860 & 0.507 & -0.000 & 0.000 & 1.860 \\
B & -0.237 & -0.216 & 1.536 & -0.018 & -0.024 & -0.014 & 1.536 \\
B & -0.256 & -0.236 & 1.474 & -0.017 & 0.023 & 0.014 & 1.474 \\
B & -0.048 & -0.860 & -1.156 & -0.155 & -0.000 & -0.000 & 1.156 \\
B & -0.792 & -0.117 & -1.155 & 0.274 & -0.000 & 0.000 & 1.155 \\
B & -1.572 & -2.454 & -1.143 & -0.518 & -0.001 & 0.000 & 1.143 \\
B & -2.680 & -1.341 & -1.139 & 0.123 & -0.000 & -0.001 & 1.139 \\
\hline
N & -0.738 & -0.463 & 5.641 & -0.275 & -0.010 & 0.004 & 5.641 \\
N & -0.771 & -0.431 & 5.639 & -0.257 & -0.001 & -0.011 & 5.639 \\
N & -0.218 & -0.768 & 2.630 & 0.003 & -0.001 & 0.004 & 2.630 \\
N & -0.629 & -0.358 & 2.625 & 0.240 & 0.003 & -0.003 & 2.625 \\
N & -0.084 & -0.173 & 1.970 & 0.077 & 0.001 & 0.000 & 1.970 \\
N & 0.171 & 0.283 & 1.317 & -0.010 & -0.031 & 0.334 & 1.359 \\
N & 0.245 & 0.206 & 1.313 & -0.053 & 0.273 & -0.193 & 1.355 \\
N & 0.136 & 0.238 & 1.180 & -0.009 & 0.028 & -0.302 & 1.218 \\
N & 0.202 & 0.167 & 1.173 & -0.048 & -0.247 & 0.174 & 1.211 \\
N & 0.174 & 0.118 & 0.979 & 0.048 & -0.267 & -0.154 & 1.026 \\
N & -0.047 & -0.153 & 0.969 & -0.074 & -0.001 & 0.001 & 0.969 \\
N & -0.191 & -0.009 & 0.968 & 0.009 & 0.001 & -0.001 & 0.968 \\
N & 0.069 & -0.007 & 0.940 & 0.066 & -0.000 & -0.000 & 0.940 \\
N & 0.145 & 0.094 & 0.883 & 0.044 & 0.244 & 0.141 & 0.927 \\
\hline
\end{tabular} 
\end{table*}

\bibliography{references}

\end{document}